\documentclass[fleqn,usenatbib]{mnras}

\usepackage{newtxtext,newtxmath}

\usepackage[T1]{fontenc}
\usepackage{ae,aecompl}

\usepackage{graphicx}	
\usepackage{amsmath}	
\usepackage{amssymb}	

\newcommand{\G}{{\it Gaia}}
\newcommand{\Teff}{$T_\mathrm{eff}$}
\newcommand{\Teffn}[1]{$T_\mathrm{eff#1}$}
\newcommand{\Logg}{$\log g$}
\newcommand{\Feh}{$\mathrm{[Fe/H]}$}
\newcommand{\Mh}{$\mathrm{[M/H]}$}

\newcommand{\TC}{\textit{The Cannon}}
\newcommand{\kms}{\,km\,s$^{-1}$}
\newcommand{\cms}{\,cm\,s$^{-2}$}
\newcommand{\vsin}{$v \sin i$}

\title[GALAH: unresolved triple Sun-like stars]{The GALAH survey: unresolved triple Sun-like stars discovered by the {\it Gaia} mission}

\author[K. \v{C}otar et al.]{
Klemen~\v{C}otar,$^{1}$\thanks{Contact e-mail: \href{mailto:klemen.cotar@fmf.uni-lj.si}{klemen.cotar@fmf.uni-lj.si}} Toma\v{z}~Zwitter,$^{1}$ Gregor~Traven,$^{2}$ Janez~Kos,$^{1}$ 
Martin~Asplund,$^{3,4}$ \newauthor
Joss~Bland-Hawthorn,$^{3,5,6}$ Sven~Buder,$^{7}$ Valentina D'Orazi,$^{8}$ Gayandhi M. De Silva,$^{9}$  \newauthor 
Jane~Lin,$^{3,4}$ Sarah~L.~Martell,$^{3,10}$ Sanjib~Sharma,$^{3,5}$ Jeffrey~D.~Simpson,$^{10}$  \newauthor 
Daniel B. Zucker,$^{9}$
Jonathan~Horner,$^{11}$ Geraint~F.~Lewis,$^{5}$ Thomas~Nordlander,$^{3,4}$ \newauthor 
Yuan-Sen~Ting,$^{12,13,14}$ Rob~A.~Wittenmyer,$^{11}$ and~the~GALAH~collaboration
\\
\\
$^{1}$ Faculty of Mathematics and Physics, University of Ljubljana, Jadranska 19, 1000 Ljubljana, Slovenia \\
$^{2}$ Lund Observatory, Department of Astronomy and Theoretical Physics, Box 43, SE-221 00 Lund, Sweden \\
$^{3}$ Center of Excellence for Astrophysics in Three Dimensions (ASTRO-3D), Australia \\
$^{4}$ Research School of Astronomy \& Astrophysics, Australian National University, ACT 2611, Australia\\
$^{5}$ Sydney Institute for Astronomy, School of Physics, A28, The University of Sydney, NSW 2006, Australia \\
$^{6}$ Miller Professor, Miller Institute, UC Berkeley, Berkeley CA 94720\\
$^{7}$ Max Planck Institute for Astronomy (MPIA), Koenigstuhl 17, D-69117 Heidelberg, Germany \\
$^{8}$ INAF Osservatorio Astronomico di Padova, vicolo dell'Osservatorio 5, I-35122, Padova, Italy \\
$^{9}$ Department of Physics and Astronomy, Macquarie University, Sydney, NSW 2109, Australia \\
$^{10}$ School of Physics, UNSW, Sydney, NSW 2052, Australia\\
$^{11}$ Centre for Astrophysics, University of Southern Queensland, Toowoomba, QLD 4350, Australia \\
$^{12}$ Institute for Advanced Study, Princeton, NJ 08540, USA\\
$^{13}$ Observatories of the Carnegie Institution of Washington, 813 Santa Barbara Street, Pasadena, CA 91101, USA\\
$^{14}$ Department of Astrophysical Sciences, Princeton University, Princeton, NJ 08544, USA
}

\date{Accepted XXX. Received YYY; in original form ZZZ}

\pubyear{2019}

\begin{document}
\label{firstpage}
\pagerange{\pageref{firstpage}--\pageref{lastpage}}
\maketitle

\begin{abstract}
	The latest {\it Gaia} data release enables us to accurately identify stars that are more luminous than would be expected on the basis of their spectral type and distance. During an investigation of the 329 best Solar twin candidates uncovered among the spectra acquired by the GALAH survey, we identified 64 such over-luminous stars. In order to investigate their exact composition, we developed a data-driven methodology that can generate a synthetic photometric signature and spectrum of a single star. By combining multiple such synthetic stars into an unresolved binary or triple system and comparing the results to the actual photometric and spectroscopic observations, we uncovered 6 definitive triple stellar system candidates and an additional 14 potential candidates whose combined spectrum mimics the Solar spectrum. Considering the volume correction factor for a magnitude limited survey, the fraction of probable unresolved triple stars with long orbital periods is $\sim$2~\%. Possible orbital configurations of the candidates were investigated using the selection and observational limits. To validate the discovered multiplicity fraction, the same procedure was used to evaluate the multiplicity fraction of other stellar types. 
\end{abstract}


\section{Introduction}
The investigation of Solar-type stars in the Solar neighbourhood has revealed that around half of them are found in binary or more complex stellar systems \citep{2010ApJS..190....1R, 2013ARA&A..51..269D, 2017ApJS..230...15M}. Of all multiple systems, about 13~\% are part of higher-order hierarchical systems \citep{2010ApJS..190....1R, 2014AJ....147...87T}. Beyond the Solar neighbourhood, the angular separation between members of such multiple star systems becomes too small for their components to be spatially resolvable in the sky. As a result, all stellar components contribute to the light observed by spectroscopic, photometric, and astrometric surveys. It has been suggested that the population of binaries in the field could be even higher than in the Solar neighbourhood \citep{2000A&A...361..770Q}, therefore a combination of multiple complementary approaches must be used to detect and analyse multiple stellar systems with different properties \citep{2017ApJS..230...15M}.

If the orbital period of such a system is relatively short, with high orbital velocities, it can be spectroscopically identified as a multiple system in two ways. When the components are of comparable luminosities, the effect of multiple absorption lines can be observed in the system's spectrum. Such an object is also known as a double-lined binary (SB2) system \citep{2004A&A...424..727P, 2010AJ....140..184M, 2017PASP..129h4201F, 2017A&A...608A..95M}. By contrast, a single-lined binary (SB1) system does not show the same effect, as the secondary component is too faint to significantly contribute to the observables. Short period SB1 systems can be identified from the periodic radial velocity variations if multi-epoch spectroscopy is available \citep{1991A&A...248..485D, 2004A&A...418..989N, 2011AJ....141..200M, 2016AJ....151...85T, 2018ApJ...854..147B}. Other extrema are very wide binaries \citep{1988ApJ...335L..47G, 1990AJ....100.1968C, 1995ApJ...441..200G, 2009ApJ...703.1511K, 2011ApJS..192....2S, 2017MNRAS.472..675A, 2018MNRAS.480.4302C}, and co-moving pairs \citep{2017AJ....153..257O, 2019AJ....157...78J} that can only be identified by their spatial proximity and common velocity vector.

\citet{2013ARA&A..51..269D} summarized that the majority of Solar-type stars are part of binary systems with periods of hundreds to thousands of years, whose period distribution reaches a maximum at $\log(P)\approx5$, for $P$ measured in days. Because of the wide separation and long orbital periods of the components in such a scenario, the radial velocity variations will have both low amplitude and long period. This makes them challenging to detect in large-scale spectroscopic surveys, which typically last for less than a decade, and have a low number of revisits. A spectrum of such a binary or triple still contains a spectroscopic signature of all members, and those contributions can be disentangled into individual components \citep{2018MNRAS.473.5043E, 2018MNRAS.476..528E}. Such a decomposition is easier when a binary consists of spectrally different stars \citep{2005A&A...440..995S, 2007MNRAS.382.1377R, 2012MNRAS.419..806R, 2013AJ....146...82R, 2016MNRAS.458.3808R}, but becomes much harder or near-impossible when the composite spectrum consists of contribution from near-identical stars whose individual radial velocities are almost identical \citep{2015IJAsB..14..173B}. In that case, additional photometric and distance measurement have to be used to constrain possible combinations \citep{2018ApJ...857..114W}. If spectroscopic data are not available, determination of multiples can be attempted purely on the basis of photometric information \citep{1997A&A...327..598F, 1998MNRAS.300..977H, 2016MNRAS.455.3009M, 2018ApJ...857..114W}, but such approaches are limited to certain stellar types, and yield results that might be polluted with young pre-main-sequence stars \citep{2018arXiv181010435Z}.

In the scope of the GALactic Archaeology with HERMES (GALAH) survey, most stars are observed at only one epoch, limiting the number of available data points that can be used to identify and characterize unresolved spectroscopic binaries. For this study, GALAH observations were complemented by multiple photometric surveys presented in Section \ref{sec:data}. Both spectroscopic (Section \ref{sec:s_model}) and photometric (Section \ref{sec:p_model}) observations were used to create data-driven models of a single star. Those models were used to analyse Solar twin candidates detected amongst the GALAH spectra (Section \ref{sec:solar_twins_sel}). Sections \ref{sec:multi_fit} and \ref{sec:single_fit} describe the employed fitting procedure that was used to determine multiplicity and physical parameters of the twin systems. In Section \ref{sec:orital_periods} we determine constraints on orbital periods of the detected triple candidates. Constraints are imposed using the observational limits and simulations described in Section \ref{sec:simulations}, where a population bias introduced by the magnitude-limited survey is assessed in Section \ref{sec:bias}. Concluding Section \ref{sec:concl} summarises the results and introduces additional approaches that could be used to verify the results.

\section{Data}
\label{sec:data}

\begin{figure}
	\centering
	\includegraphics[width=\columnwidth]{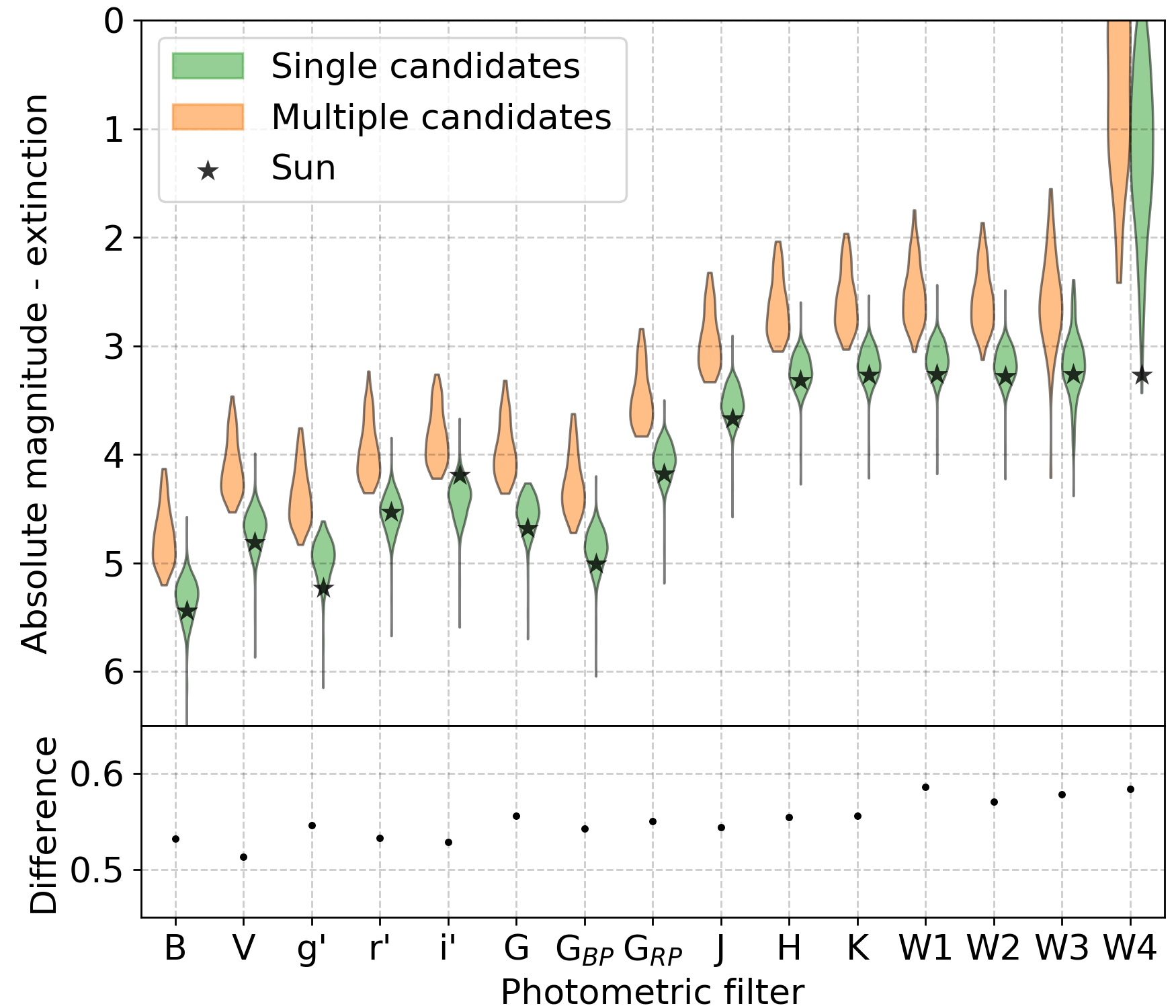}
	\caption{Violin density plots showing the distribution of extinction-corrected absolute magnitudes in multiple photometric systems. Separate distributions are given for stars that we considered as single and multiple in our analysis. Star symbols indicate absolute magnitude of the Sun \citep{2018ApJS..236...47W}. The bottom panel shows a difference between the median magnitudes of both distributions.}
	\label{fig:viol_photometry}
\end{figure}

In this work, we analyse a set of high-resolution stellar spectra that were acquired by the High Efficiency and Resolution Multi-Element Spectrograph \citep[HERMES,][]{2010SPIE.7735E..09B, 2015JATIS...1c5002S}, a multi-fibre spectrograph mounted on the $3.9$-metre Anglo-Australian Telescope (AAT). The spectrograph has a resolving power of R $\sim 28,000$ and covers four wavelength ranges (4713 -- 4903 \AA, 5648 -- 5873 \AA, 6478 -- 6737 \AA, and 7585 -- 7887 \AA), frequently also referred to as spectral arms. Observations used in this study have been taken from multiple observing programmes that make use of HERMES spectrograph: the main GALAH survey \citep{2015MNRAS.449.2604D}, the K2-HERMES survey \citep{2018AJ....155...84W}, and the TESS-HERMES survey \citep{2018MNRAS.473.2004S}. Those observing programmes mostly observe stars at higher Galactic latitudes ($|b|$>$10^\circ$), employ different selection functions, but share the same observing procedures, reduction, and analysis pipeline \citep[internal version $5.3$,][]{2017MNRAS.464.1259K}. All three programmes are magnitude-limited, with no colour cuts (except the K2-HERMES survey), and observations predominantly fall in the V magnitude range between 12 and 14. This leads to an unbiased sample of mostly southern stars that can be used for different population studies, such as multiple stellar systems in our case. Additionally, stellar atmospheric parameters and individual abundances derived from spectra acquired during different observing programmes are analysed with the same \TC\ procedure \citep[internal version 182112 that uses parallax information to infer \Logg\ of a star,][]{2015ApJ...808...16N, buder2018}, so they are inter-comparable. \TC\ is a data-driven interpolation approach trained on a set of stellar spectra that span the majority of the stellar parameter space \citep[for details, see][]{buder2018}. Whenever we refer to valid or unflagged stellar parameters in the text, only stars with the quality flag \texttt{flag\_cannon} equal to 0 were selected.

With the latest \G\ DR2 release \citep{2016A&A...595A...1G, 2018arXiv180409365G}, the determination of distance for stars within a few kpc away from the Sun becomes straightforward, and the derived distances are accurate to a few percent. Along with the measurements of parallax and proper motion, stellar magnitudes in up to three photometric bands (G, G$_{BP}$ and G$_{RP}$) are provided. Coupling those two measurements together, we can determine the absolute magnitude and luminosity of a star. This can be done for the majority of stars observed in the scope of the GALAH survey as more than 99~\% of them can be matched with sources in \G\ DR2. Although the uncertainty in \G\ mean radial velocities is much larger than those from GALAH \citep{2018arXiv180406344Z}, especially at its faint limit, they can still be used in the multi-epoch analysis (Section \ref{sec:orbits_rv}) to determine the lower boundary of its variability.

Because observations of both surveys were acquired at approximately the same epoch ranges, wide binaries with long orbital periods would show little variability in their projected velocities. Observational time-series were therefore extended into the past with the latest data release from The Radial Velocity Experiment \citep[RAVE DR5,][]{2017AJ....153...75K} that also surveyed the southern part of the sky. More recent spectra of accessible (stars' declination >~$-25^{\circ}$ and G magnitude <~$13.5$) multiple candidates were acquired by the high-resolution Echelle spectrograph (with the resolving power R $\sim$20,000) mounted on the $1.82$-m Copernico telescope located at Cima Ekar in Asiago, Italy. As this telescope is located in a northern observatory, we were able to observe only a limited number of stars. Every observation consisted of two half-hour long exposures that were fully reduced, shifted to the heliocentric frame, combined together, and normalized. The system's radial velocity was determined by cross-correlating the observed spectrum with that of the Sun. Data from those two additional spectroscopic surveys enabled us to further constrain the variability of the analysed systems.

For a broader wavelength coverage, additional photometric data were taken from three large all-sky surveys. In the visual part of the spectrum we rely on the AAVSO Photometric All-Sky Survey \citep[APASS,][]{2015AAS...22533616H} B, V, g', r', i' bands that are supplemented by the Two Micron All-Sky Survey \citep[2MASS,][]{2006AJ....131.1163S} J, H, K$_S$ bands, and the Wide-field Infrared Survey Explorer \citep[WISE,][]{2010AJ....140.1868W} W1, W2 bands. All of those surveys were matched with the GALAH targets, which resulted in up to 13 photometric observations per star. Photometric values and their uncertainties were taken as published in these catalogues, ignoring any specific quality flags. During the initial investigation of their usefulness, WISE W3, and W4 bands proved to be unreliable for our application and were therefore removed from further use. The main reason for their removal is a large scatter in magnitude measurements of similar stars and a strong overlap between single and multiple stars evident in Figure \ref{fig:viol_photometry}.

\begin{figure}
	\centering
	\includegraphics[width=\columnwidth]{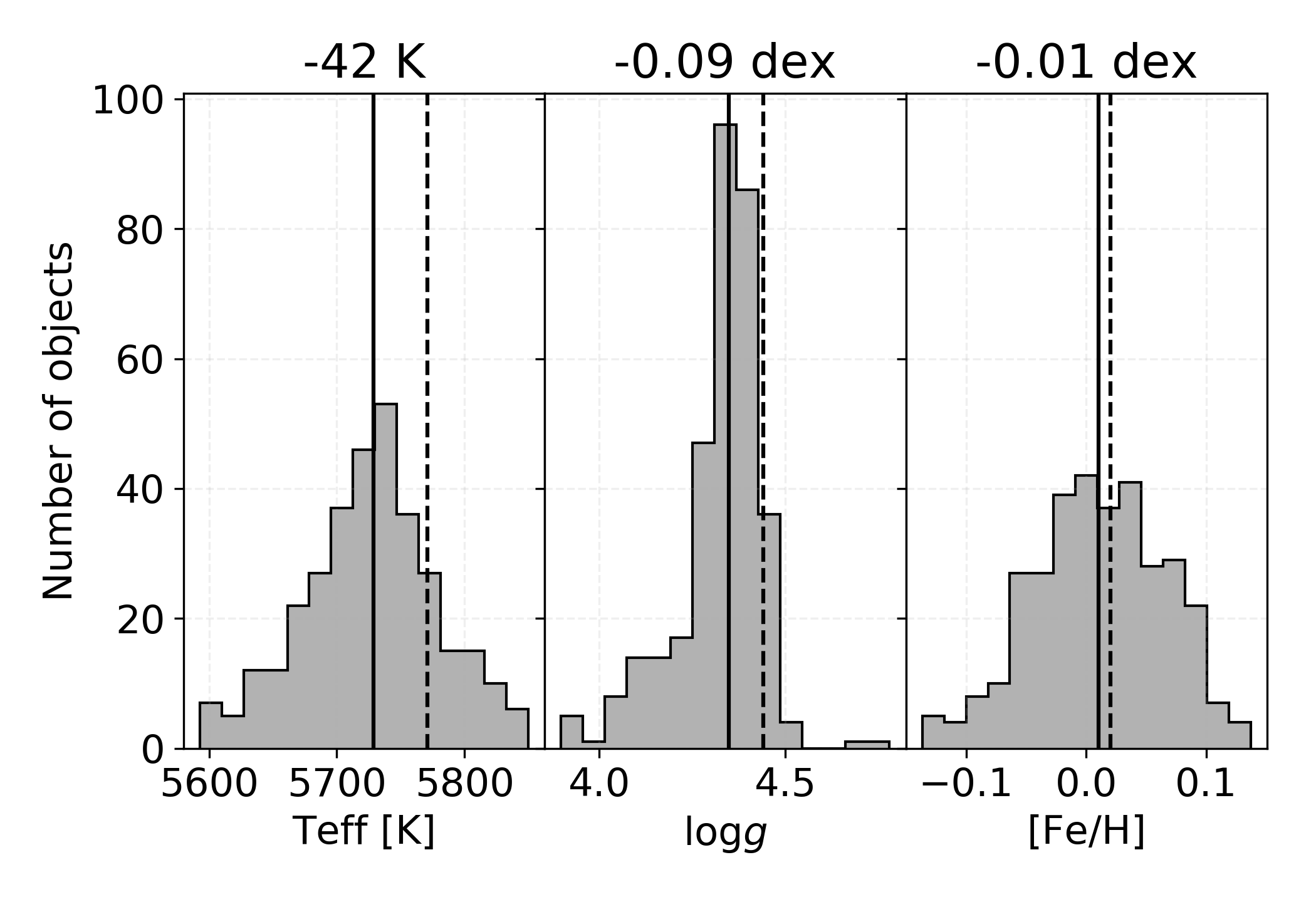}
	\caption{Distribution of physical parameters for discovered Solar twin candidates. Values above the plots represent difference between median of the distribution (solid vertical line) and actual Solar values (dashed vertical line).}
	\label{fig:twins_stats}
\end{figure}

\section{Solar-like spectra}
\label{sec:solar_twins_sel}
The selection of the most probable Solar twin candidates observed in the scope of the GALAH survey was performed purely on the basis of spectral comparison using the Canberra distance metric \citep{Lance1967MixedDataCP}. It is defined as
\begin{equation}
\label{equ:equ_canberra}
distance_{Canberra}(f_{\sun}, f_{obs}) = \sum_{\lambda}^{\ } \frac{|f_{\sun,\lambda} - f_{obs,\lambda}|}{|f_{\sun,\lambda}| + |f_{obs,\lambda}|},
\end{equation}
where $f_{\sun}$ is the reference Solar spectrum and $f_{obs}$ observed spectrum. This avoided the need for prior knowledge about the physical parameters of the considered stars. Observed spectra were compared with reference twilight Solar spectrum that was acquired by the same spectrograph. The comparison was done in two steps. First, the observed spectra were compared arm by arm, where only 7~\% of the best matching candidates were selected for every HERMES arm. The selection does not consist of only one hard threshold on spectral similarity, but it varies as a power law of spectral SNR. Larger discrepancies between spectra are therefore allowed for spectra with low SNR. The final selection consisted of a set of stars whose spectra were adequately similar to the Solar spectrum in every arm. Further processing consisted of modelling the spectral noise and the comparison of individual chemical abundances. For details about spectral comparison and Solar twin determination see \v{C}otar et al. (in preparation). The parameter distribution for the selected set of objects is presented in Figure \ref{fig:twins_stats}. As the computation of \Logg\ uses prior knowledge about the absolute magnitude of a star, lower \Logg\ values are attributed to objects that exhibit excess luminosity. 

\begin{figure}
	\centering
	\includegraphics[width=\columnwidth]{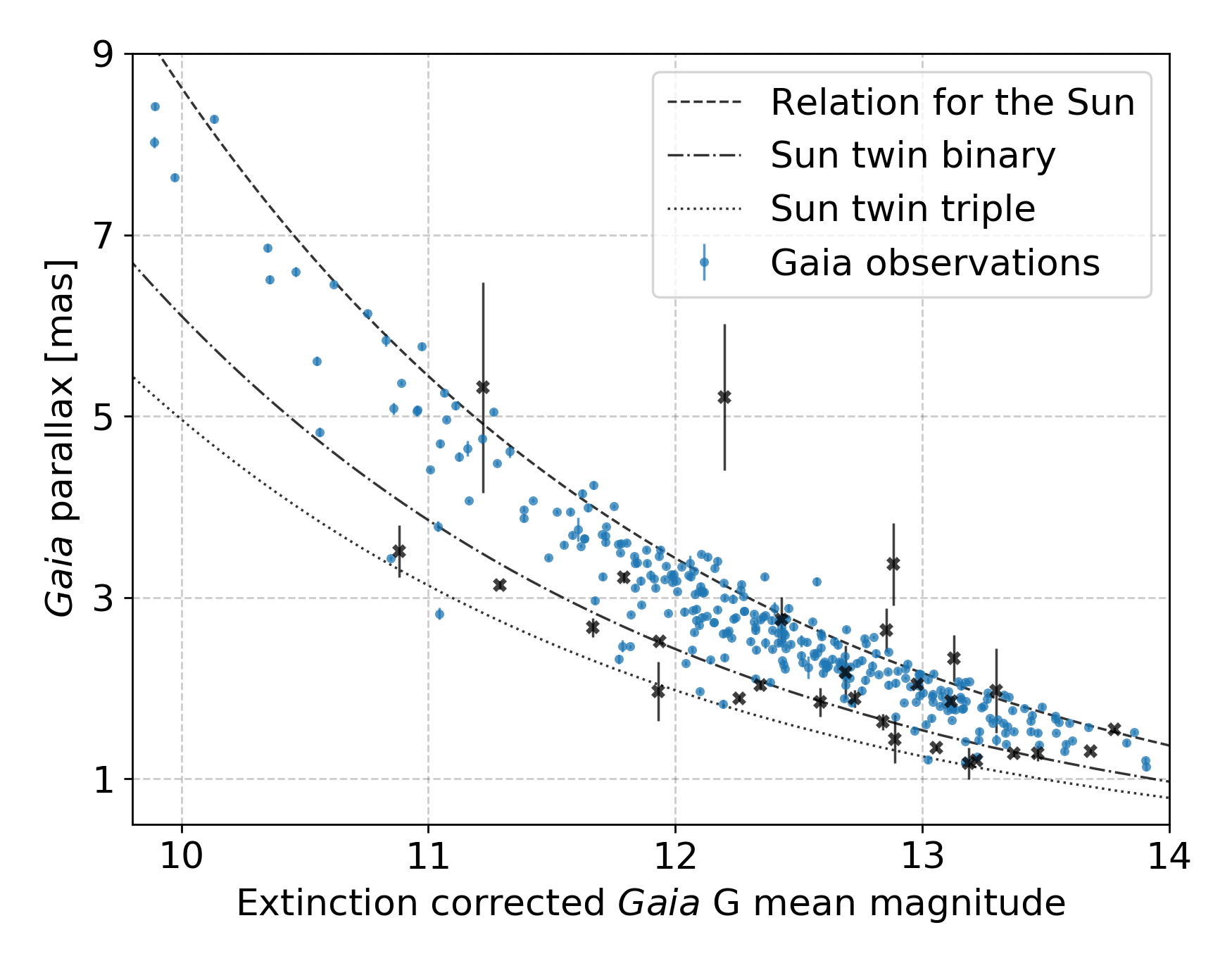}
	\caption{Parallax versus measured apparent \G\ G magnitude for our Solar twin candidates. Stars marked with black crosses have large normalised astrometric uncertainties (RUWE~>~$1.4$) which may lead to wrongly determined distance and consequently multiplicity results. The dashed line represents a theoretical absolute magnitude of the Sun as it would be observed at different parallaxes. Similarly, the relation for a binary and a triple system composed of multiple Solar twins are plotted with the dash-dotted and dotted line.}
	\label{fig:par_gmean}
\end{figure}

\subsection{Candidate multiple systems}
\label{sec:multi_cand}
Among the determined Solar twin candidates, we noticed a photometric trend that is inconsistent with the distance of an object that resembles the Sun. Solar twins, mimicking the observed Solar spectrum, should also be similar to it in all other observables such as luminosity, effective temperature, surface gravity, chemical composition and absolute magnitude. Plotting their apparent magnitude against \G\ parallax measurement (Figure \ref{fig:par_gmean}), all of the detected stars should lie near or on the theoretical line, describing the same relation for the Sun observed at different parallaxes. As the magnitude of the Sun is not directly measured by the \G\ or determined by the \G\ team, we computed its absolute magnitude using the relations published by \citet{2018arXiv180409368E} that connect the \G\ photometric system with other photometric systems. The reference Solar magnitudes (in multiple filters) that were used in the computation were taken from \citet{2018ApJS..236...47W}. The resulting absolute G magnitude of the Sun is $4.68 \pm 0.02$, where the uncertainty comes from the use of multiple relations. This value also coincides with the synthetic \G\ photometry produced by \citet{2018MNRAS.479L.102C}, who determined magnitude of the Sun to be M$_{G, \sun} = 4.67$.

\begin{figure}
	\centering
	\includegraphics[width=\columnwidth]{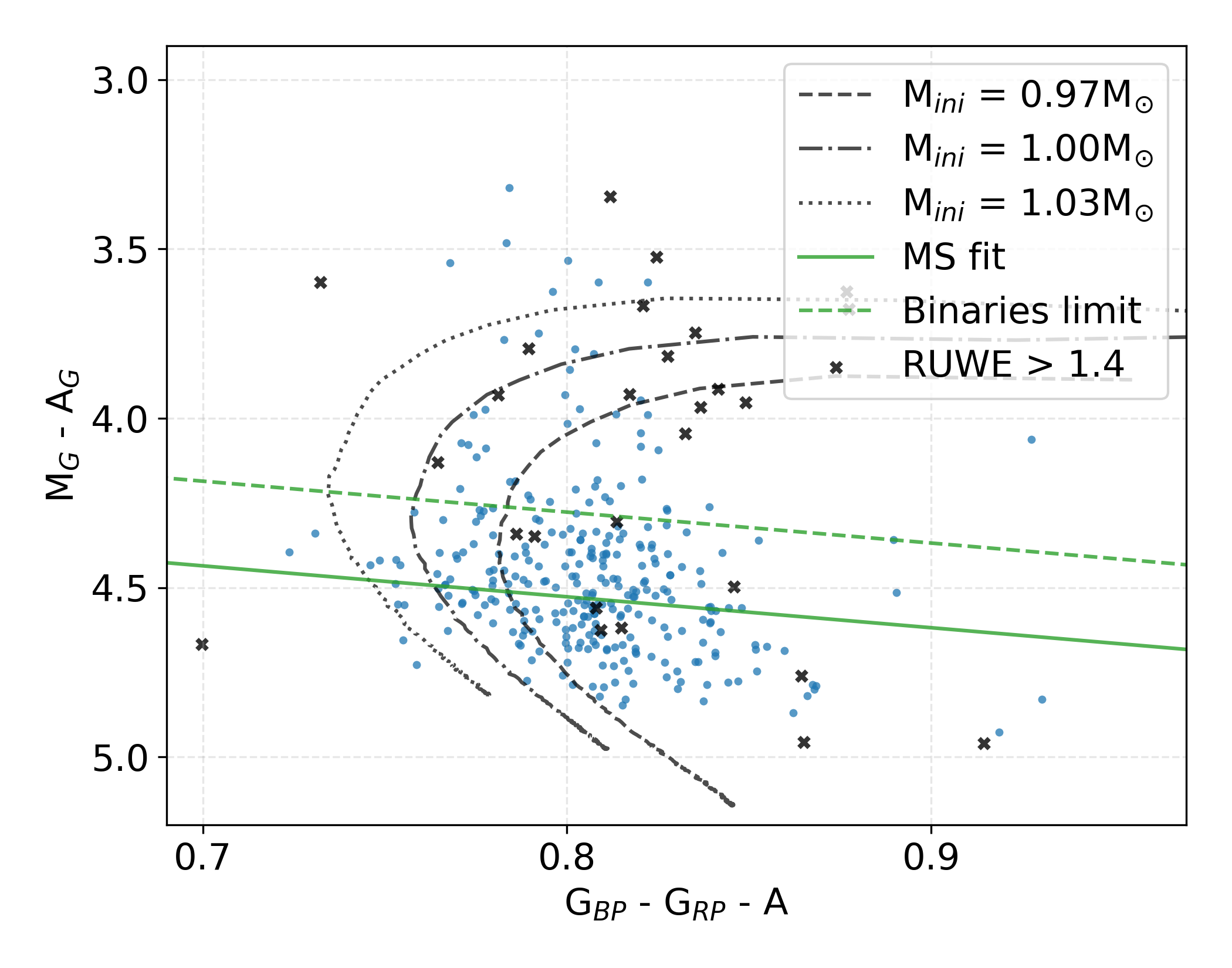}
	\caption{\G\ extinction corrected absolute G magnitude and colour index computed from G$_{BP}$, and G$_{RP}$ bands. Stars marked with black crosses have large normalised astrometric uncertainty (RUWE~>~$1.4$). The green dashed line represents a threshold that was used as a delimiter between objects treated as multiple and single stars. Overlaid evolutionary tracks, constructed from the PARSEC isochrones \citep{2017ApJ...835...77M}, represent an evolution of stars with Solar-like initial mass M$_{ini}$ and metallicity \Mh~=~0 for stellar ages between $0.1$ and $12$~Gyr.}
	\label{fig:gabs_colour}
\end{figure}

Within our sample of the probable Solar twins, we identified 64 stars that show signs of being too bright at a given parallax. In Figure \ref{fig:par_gmean} they are noticeable as a sequence of data points that lie below the theoretical line and are parallel to it. Another even more obvious indication of their excess luminosity is given by the colour-magnitude diagram in Figure \ref{fig:gabs_colour}. There, the same group of stars is brighter by $\sim$0.7 magnitude when comparing stars with the same colour index. As both groups of stars are visually separable, the multiple stellar candidates can easily be isolated by selecting objects with extinction corrected absolute G magnitude above the binary limit line shown in Figure \ref{fig:gabs_colour}. To compute the absolute magnitudes, we used the distance to stars inferred by the Bayesian approach that takes into account the distribution of stars in the Galaxy \citep{2018AJ....156...58B}. As the reddening published along the \G\ DR2 \citep{2018A&A...616A...8A} could be wrong for stars located away from the used set of isochrones, we took the information about the reddening at specific sky locations and distances from the all-sky three-dimensional dust map produced by \citet{2017A&A...606A..65C}. To infer a band dependent extinction from the acquired reddening, a reddening coefficient ($R$) was used. The values of $R$, considering the extinction law $R_V = 3.1$, were taken from the tabulated results published in \citet{2011ApJ...737..103S}. 

To determine the limiting threshold between single and multiple candidates, we first fit a linear representation of the main sequence to the median of the absolute magnitudes distribution in the $0.02$~mag wide colour bins. The lower limit for the binaries was placed $0.25$~mag above the fitted line.

Confirmation that this extra flux could be contributed by the unseen companion also comes from other photometric systems, where the distribution of absolute magnitudes for both groups is shown in Figure \ref{fig:viol_photometry}. On average, multiple candidates are brighter by $\sim$0.55 magnitude in every considered band. For an identical binary system, a measured magnitude excess would be $0.75$, and $1.2$ for a triple system. As the observed difference is not constant in every band, as would be expected for a system composed of identical stars, we expect some differences in parameters between the components of the system. This can be said under the assumption that all considered photometric measurements were performed at the same time. Of course, that is not exactly true in our case as the acquisition time between different photometric surveys can vary by a few to 10 years. As the Sun-like stars are normal, low activity stars, this effect is most probably negligible, but events like occultations between the stars in the system can still occur. 

\section{Single star models}
\label{sec:models_all}
Once the selection of interesting stars was performed, we began with the analysis of their possible multiplicity. Our procedure for the analysis of suspected multiple stellar systems is based on spectroscopic and photometric data-driven single star models that were constructed from observations taken from multiple large sky surveys. With this approach, we exclude assumptions about stellar properties and populations that are usually used to generate synthetic data. In this section, we describe approaches that were used to create those models.

\subsection{Spectroscopic model}
\label{sec:s_model}
\begin{figure}
	\centering
	\includegraphics[width=\columnwidth]{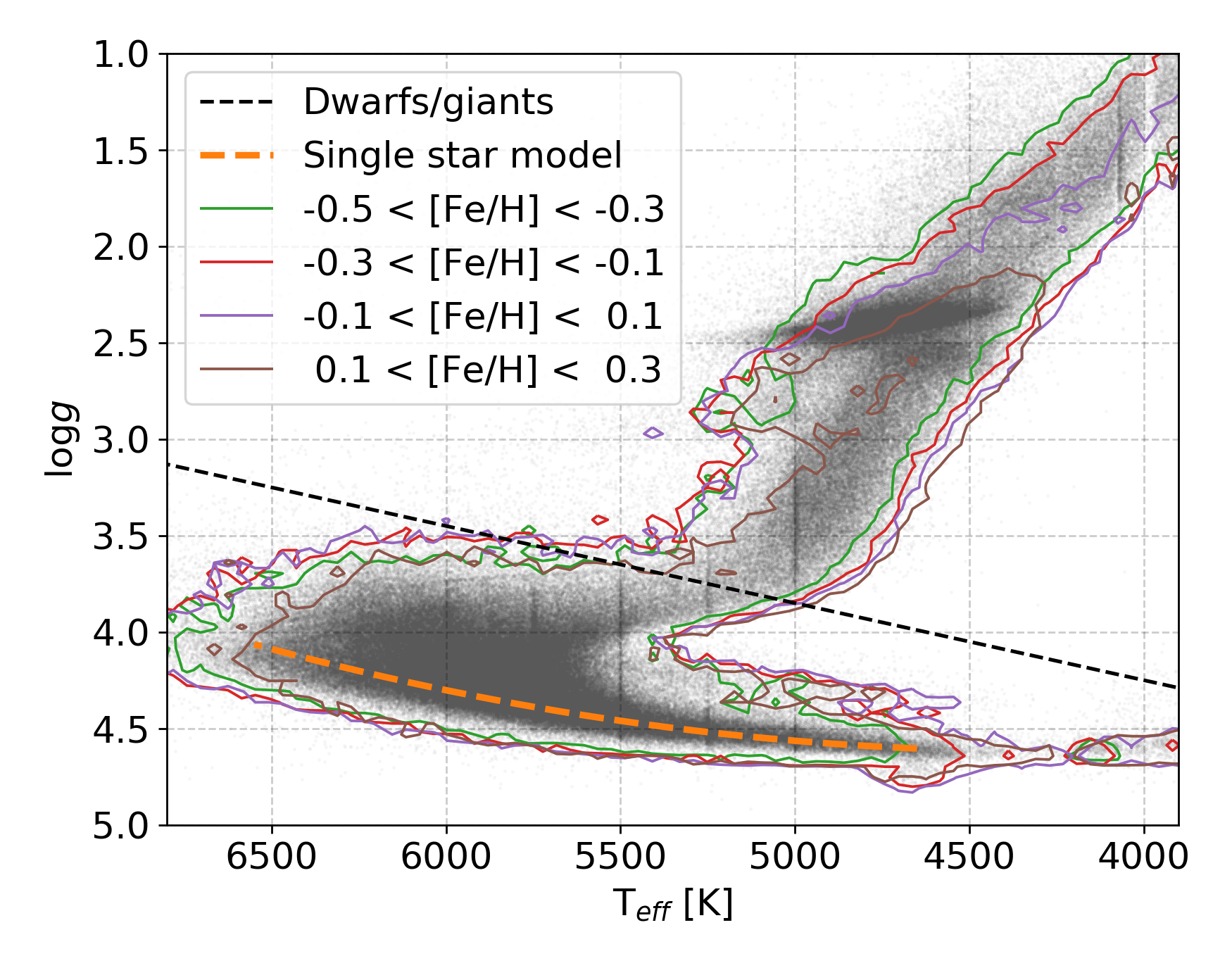}
	\caption{Complete observational set of valid stellar parameters shown as a varied density of grey dots. Contours around the diagram illustrate a coverage of parameter space in different \Feh\ bins. Overlaid orange dashed curve represents a relation on the main sequence from where single stars considered in the fit are drawn. Additionally, black dashed linear line represents arbitrarily defined limit between giant and dwarf stars that were used in the process of training a spectroscopic single star model.}
	\label{fig:kiel_cannon}
\end{figure}

Every stellar spectrum can be largely described using four basic physical stellar parameters: \Teff, \Logg, chemical composition, and \vsin. To construct a model that would be able to recreate a spectrum corresponding to any conceivable combination of those parameters, we used a data-driven approach named \TC. The model was trained on a set of normalised GALAH spectra that meet the following criteria: the spectrum must not be flagged as peculiar \citep{2017ApJS..228...24T}, have a signal to noise ratio (SNR) per resolution element in the green arm >~20, does not contain any monitored reduction problems and have valid \TC\ stellar parameters. Additionally, we limited our set to main sequence dwarf stars (below the arbitrarily defined line shown in Figure \ref{fig:kiel_cannon}) as giants are not considered for our analysis. Additionally, the decision not to consider giants was taken as a result of the fact that accurate modelling of their spectra requires information about their luminosity. It should be noted that the application of these limits does not ensure that our training set is completely free from spectra of unresolved (or even clearly resolved SB2 binaries), as would be desired in the case of an ideal training set.

In order to train the model, all spectra were first shifted to the rest frame by the reduction pipeline \citep{2017MNRAS.464.1259K}, and then linearly interpolated onto a common wavelength grid. The training procedure consists of minimising a loss function between an internal model of \TC\ and observations for every pixel of a spectrum \citep{2015ApJ...808...16N}.

The result of this training procedure are quadratic relations that take desired stellar parameters \Teff, \Logg, \Feh, and \vsin\ to reconstruct a target spectrum. Spectra generated in this manner are trustworthy only within the parameter space defined by the training set, where the main limitation is the effective temperature which ranges from $\sim$4600 to $\sim$6700~K on the main sequence. Spectra of hotter stars are easy to reproduce, but they lack elemental absorption lines that we would like to analyse. On the other hand, spectra of colder stars are packed with molecular absorption lines and therefore harder to reproduce and analyse. The model itself can be used to extrapolate spectra outside the initial training set, but as they can not be verified, they were not considered to be useful for the analysis.

\subsection{Photometric model}
\label{sec:p_model}
With the use of a model that produces normalised spectra for every given set of stellar parameters, we lose all information about the stars' colour, luminosity, and spectral energy distribution. This can be overcome using another model that generates the photometric signature of a desired star. To create this kind of a model, we first collected up to 13 apparent magnitudes from the selected photometric surveys (\G, APASS, 2MASS, and WISE) for every star in the GALAH survey. Whenever possible, these values were converted to absolute magnitudes using the distance to stars inferred by the Bayesian approach \citep{2018AJ....156...58B}. Before using the pre-computed published distances, we removed all sources whose computed astrometric re-normalised unit-weight error \citep[RUWE,][]{ruwe} was greater than $1.4$. The magnitudes of every individual star were also corrected for the reddening effect, except for the WISE photometric bands W1 and W2 that were considered to be extinction free.

Using the valid \TC\ stellar parameters, the inferred and corrected absolute magnitudes in multiple photometric bands were grouped into bins that contain stars within $\Delta$\Teff~=~$\pm$80~K, $\Delta$\Logg~=~$\pm$0.05~dex, and $\Delta$\Feh~=~$\pm$0.1~dex around the bin centre. As spectroscopically unresolved multiple stellar systems could still be present in this bin, median photometric values are computed per bin to minimise their effect. Extrapolation outside this grid, that covers the complete observational stellar parameter space, is not desired nor implemented as it may produce erroneous values. When a photometric signature of a star with parameters between the grid points is requested, it is recovered by linear interpolation between the neighbouring grid points. The median photometric signature for the requested stellar parameters could also be computed on-the-fly using the same binning, but we found out that this produced insignificant difference and increased the processing time by more than a factor of 2.

\subsection{Limitations in the parameter space}
As already emphasized, spectroscopic and photometric models were built on real observations and are therefore limited by the training set coverage. The limitations are visually illustrated by Figure \ref{fig:kiel_cannon}, where the dashed curve, from which single stars are drawn, is plotted over the observations. This arbitrary quadratic function is defined as:
\begin{equation}
\label{equ:logg_teff}
\log g = 2.576 + 9.48 \cdot 10^{-4} \ T_\mathrm{eff} - 1.10 \cdot 10^{-7} \ T_\mathrm{eff}^2,
\end{equation}
where values of \Teff\ and \Logg\ are given in units of K and \cms\ respectively. The polynomial coefficients were determined by fitting a quadratic function to manually defined points that represent regions with the highest density of stars on the shown Kiel diagram. From Equation \ref{equ:logg_teff}, it follows that the \Logg\ of a selected single star is not varied freely, but computed from the selected \Teff\ whose range is limited within the values 6550~\textgreater~\Teff~\textgreater~4650~K.

Focusing on Solar-like stars gives us an advantage in their modelling as the whole observational diagram in Figure \ref{fig:kiel_cannon} is sufficiently populated with stars of Solar-like iron abundance. When going towards more extreme \Feh\ values (high or low), coverage of the main sequence starts to decrease. For cooler stars, this happens at low iron abundance (\Feh~\textless~$-0.3$) and for hot stars at high abundance (\Feh~\textgreater~$0.3$). Those limits pose no problems for our analysis, unless the wrong stellar configuration is used to describe the observations. At that point, the spectroscopic fitting procedure would try to compensate for too deep or too shallow spectral lines (effect of wrongly selected \Teff) by decreasing or increasing \Feh\ beyond values reasonable for Solar-like objects.

\section{Characterization of multiple system candidates}
\label{sec:multi_fit}
For the detailed characterization of Solar twin candidates that show excess luminosity, we used their complete available photometric and spectroscopic information. The excess luminosity can only be explained by the presence of an additional stellar component or a star that is hotter or larger than the Sun. Both of those cases can be investigated and confirmed by the data and models described in Sections \ref{sec:data} and \ref{sec:models_all}. In the scope of our comparative methodology, we constructed a broad collection of synthetic single, double, and triple stellar systems that were compared and fitted to the observations.

As the measured \G\ DR2 parallaxes, and therefore inferred distances, of some objects are badly fitted or highly uncertain, the distance results provided by \citep{2018AJ....156...58B} actually yield three distinct distance estimates - the mode of an inferred distance distribution (\texttt{r\_est}) and a near and distant distance (\texttt{r\_lo}, and \texttt{r\_hi}) estimation, between which 68~\% of the distance estimations are distributed. As the actual shape of the distribution is not known and could be highly skewed, we did not draw multiple possible distances from the distribution, but only used its mode value.

\subsection{Fitting procedure}
A complete characterization and exploration fitting procedure for every stellar configuration (single, binary, and triple) consists of four consecutive steps that are detailed in the following sections. As we are investigating Solar twin spectra, the initial assumption for the iron abundance of the system is set to \Feh~=~$0$. This also includes the assumption that stars in a system are of the same age, at similar evolutionary stages, and were formed from a similarly enriched material. If that is true, we can set iron abundance to be equal for all stars in the system. This notion is supported by the simulations \citep{2019arXiv190210719K} and studies \citep{2019arXiv190402159K} of field stars showing that close stars are very likely to be co-natal if their velocity separation is small.

The observed systems must be composed of multiple main sequence stars, otherwise the giant companion would dominate the observables, and the system would not be a spectroscopic match to the Sun. Therefore their parameters \Teff\ and \Logg\ are drawn from the middle of the main sequence determined by \TC\ parameters in the scope of the GALAH survey. The Kiel diagram of the stars with valid parameters and model of the main sequence isochrone used in the fitting procedure are shown in Figure \ref{fig:kiel_cannon}.

\subsection{Photometric fitting - first step}
\label{sec:photo_fit}
With those initial assumptions in mind, we begin with the construction of the photometric signature of the selected stellar configuration. To find the best model that describes the observations, we employ a Bayesian MCMC fitting approach \citep{2013PASP..125..306F}, where the varied parameter is the effective temperature of the components. The selected \Teff\ values, and inferred \Logg\ (Equation \ref{equ:logg_teff}), are fed to the photometric model (Section \ref{sec:p_model}) to predict a photometric signature of an individual component. Multiple stellar signatures are combined together into a single unresolved stellar source using the following equation:
\begin{equation}
	M_{model} = -2.5 \log{}_{10} \Big( \sum_{i=1}^{n_s} 10^{-0.4 M_i} \Big); \ n_s=[1, 2, 3],
\end{equation}
where $M_i$ denotes absolute magnitude of a star in one of the used photometric bands, and $n_s$ number of components in a system. The newly constructed photometric signature at selected \Teff\ values is compared to the observations using the photometric log-likelihood function $\ln p_{P}$ defined as:
\begin{equation}
\label{equ:lnp_p}
	\ln p_{P}(T_\mathrm{eff} | M, \sigma) = -\frac{1}{2} \sum_{i=1}^{n_p} \Big[ \frac{(M_i - M_{model, i})^2}{\sigma_i^2} +ln(2\pi\sigma_i^2) \Big],
\end{equation}
where $M$ and $\sigma$ represent extinction corrected absolute magnitudes, and their measured uncertainties that were taken for multiple published catalogues presented in Section \ref{sec:data}. The constructed photometric model of a multiple system is represented by the variable $M_{model}$ and the number of photometric bands by $n_p$. The maximum, and most common value for $n_p$ is 13, but in some cases, it can drop to as low as 8. The MCMC procedure is employed to maximise this log-likelihood and find the best fitting stellar components.

\begin{figure*}
	\centering
	\includegraphics[width=0.9\textwidth]{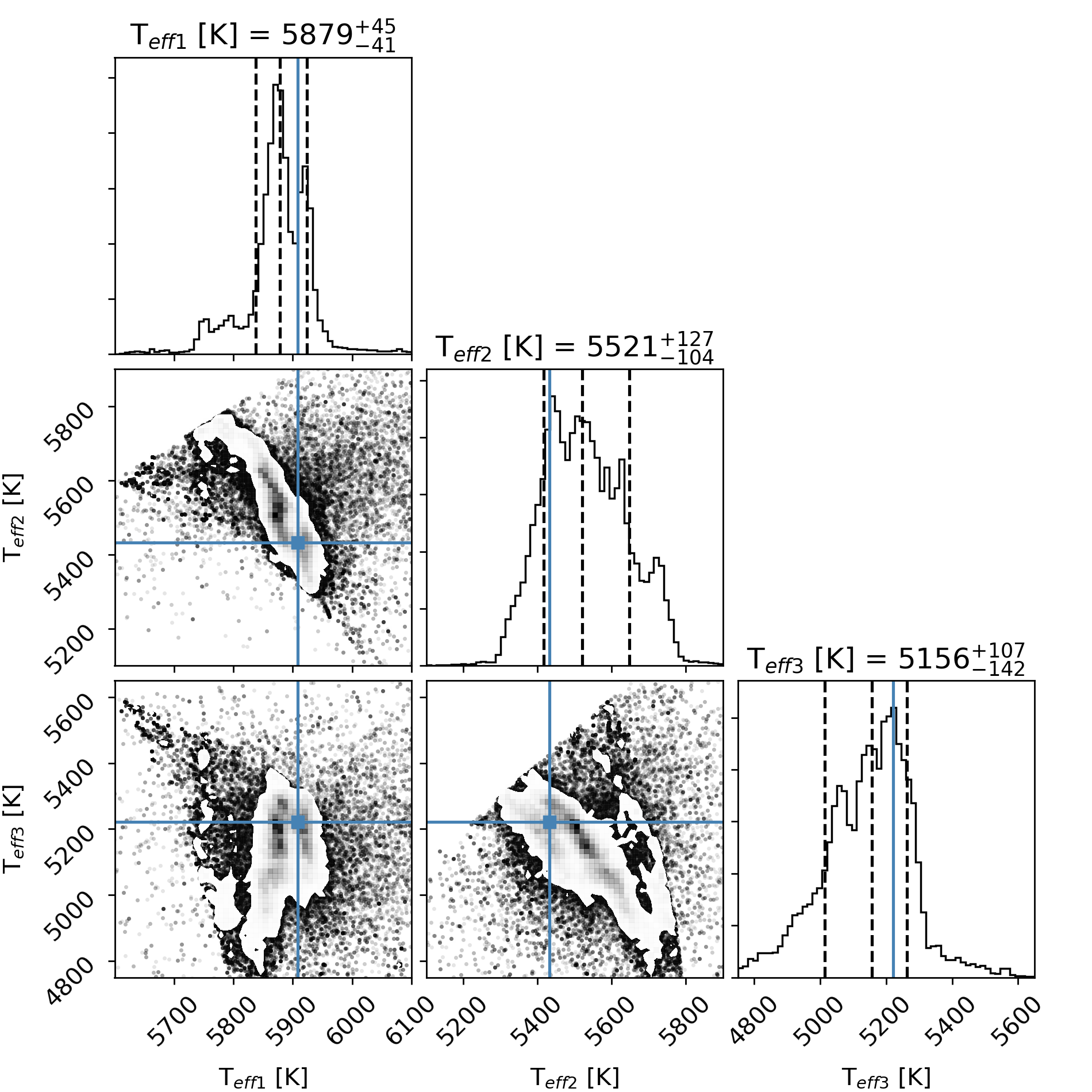}
	\caption{Distribution of considered posteriors during the initial MCMC photometric fit for one of the objects that was at the end classified as a triple candidate. As the plots show the first step in the fitting procedure that explores the complete parameter space, the distributions are not expected to be smooth because of possible multiple local minima. Values indicated on the scatter plots represent medians for the 10~\% of the best fitting solutions.}
	\label{fig:posterior_dist}
\end{figure*}

To determine the best possible combination of \Teff\ values, we initiate the fit with $1200$ uniformly distributed random combinations of initial temperature values that span the parameter space shown in Figure \ref{fig:kiel_cannon}. The number of initial combinations is intentionally high in order to sufficiently explore the temperature space. Excessive or repeated variations of initial parameters are rearranged by a prior limitation that the temperatures of components must be decreasing, therefore \Teffn{1}~>=~\Teffn{2}~>=~\Teffn{3} in the case of a triple system (example of used initial walker parameters is shown in Figure \ref{fig:teff_initial_walkers}). The initial conditions are run for 200 steps. The number of steps was selected in such a way to ensure a convergence of all considered cases (example of the walkers convergence is shown in Figure \ref{fig:walkers_logprob}). The distribution of priors for such a run is shown as a corner plot in Figure \ref{fig:posterior_dist}. After that, only the best 150 walkers are kept, their values perturbed by 2~\%, and run for another 200 steps to determine the posterior distribution of the parameters varied during the MCMC fit. This two-step run is needed to speed up the process and discard solutions with lower $\ln p_{P}$. During the initial tests, we found that the investigated parameter space can have multiple local minima which attract walkers, especially in the case of a triple system.

After the completion of all MCMC steps, the considered parameter combinations are ordered by their log-likelihood in descending order. Of those, only 10~\% of the best fitting combinations are used to compute final \Teffn{[1-3]}\ values. They are computed as the median of the selected best combinations. 

\subsection{Spectroscopic fitting - second step}
\label{sec:spectrum_fit}
After an effective temperature of the components has been determined, we proceed with the evaluation of how well they reproduce the observed spectrum. As a majority of the fitted systems do not consist of multiple components with a \Teff\ equal to Solar, the \Feh\ of the system must be slightly changed to equalize absorption strength of the simulated and observed spectral lines.

To determine the \Feh\ of the system, we first compute a simulated spectrum for every component using a spectroscopic model described in Section \ref{sec:s_model}. Individual spectra are afterwards combined using Equation \ref{equ:flux12} in the case of a binary system or Equation \ref{equ:flux123} in the case of a triple system. 

\begin{equation}
\label{equ:flux12}
\begin{aligned}  
r_{12} &= \frac{L_{2,\lambda}}{L_{1,\lambda}}; \ \lambda=[1, 2, 3, 4]\\ 
f_{model,\lambda} &= \frac{f_{1,\lambda}}{1+r_{12}} + \frac{f_{2,\lambda}}{1+1/r_{12}}
\end{aligned}
\end{equation}

\begin{equation}
\label{equ:flux123}
\begin{aligned}
r_{12} &= \frac{L_{2,\lambda}}{L_{1,\lambda}},\ \ r_{13} = \frac{L_{3,\lambda}}{L_{1,\lambda}}\\ 
f_{model,\lambda} &= \frac{f_{1,\lambda}}{1 + r_{12} + r_{13}} + \frac{r_{12} f_{2,\lambda}}{1 + r_{12} + r_{13}} + \frac{r_{13} f_{3,\lambda}}{1 + r_{12} + r_{13}}
\end{aligned}
\end{equation}

Individual normalised spectra, denoted by $f_{n,\lambda}$ in Equations \ref{equ:flux12} and \ref{equ:flux123}, are weighted by the luminosity ratios between the components ($r_{xy}$) and then summed together. 

As the HERMES spectrum covers four spectral ranges, whose distribution of spectral energy depends on stellar \Teff, different luminosity ratios also have to be used for every spectral arm. Of all of the used photometric systems, APASS filters B, g', r', and i' have the best spectral match with blue, green, red, and infrared HERMES arm. The modelled APASS magnitudes of the same stars are used to compute luminosity ratios between them.

The described summation of the spectra introduces an additional assumption about the analysed object. With this step, we assume that components have a negligible internal spread of projected radial velocities that could otherwise introduce asymmetries in the shape of observed spectral lines. The assumption allows us to combine individual components without any wavelength corrections.

Similarly, as in the previous case, a Bayesian MCMC fitting procedure was used to maximise the spectroscopic \hbox{log-likelihood} $\ln p_{S}$ defined as:
\begin{equation}
\label{equ:lnp_s}
	\ln p_{S}(\mathrm{[Fe/H]} | f, \sigma) = -\frac{1}{2} \sum_{\lambda}^{} \Big[ \frac{(f_{\lambda} - f_{model, \lambda})^2}{\sigma_{\lambda}^2} +ln(2\pi\sigma_{\lambda}^2) \Big],
\end{equation}
where $f$ and $\sigma$ represent the observed spectrum and its per-pixel uncertainty, respectively. A modelled spectrum of the system, at selected \Feh, is represented by the variable $f_{model}$ and the number of wavelength pixels in that model by $\lambda$. Combined, all four spectral bands consist of almost $16,000$ pixels. \Teff\ values of the components are fixed for all considered cases.

The MCMC fit is initiated with 150 randomly selected \Feh\ values, whose uniform distribution is centred at the initial \Feh\ value of the system and has a span of $0.4$ dex. All of the initiated walkers are run for 100 steps. At every \Feh\ level, a new simulated spectrum composite is generated and compared to the observed spectrum by computing log-likelihood $\ln p_{S}$ of a selected \Feh\ value. The range of possible \Feh\ values considered in the fit is limited by a flat prior between $-0.5$ and $0.4$. 

By the definition of \Feh\ in the scope of GALAH \TC\ analysis, the parameter describes stellar iron abundance and not its metallicity as commonly used in the literature. Therefore only spectral absorption regions of un-blended Fe atomic lines are used to compute the spectral log-likelihood. Having to fit only one variable at a time, the solution is easily found and computed as a median value of all posteriors considered in the fit.

\subsection{Final fit - third step}
\label{sec:final_fit}
A changed value of \Feh\ for the system will introduce subtle changes to its photometric signature, therefore we re-initiate the photometric fitting procedure. It is equivalent to the procedure described in Section \ref{sec:spectrum_fit}, but with much narrower initial conditions. These new initial conditions are uniformly drawn from the distribution centred at \Teff\ values determined in the first step of the fitting procedure. The width of the uniform distribution is equal to 100~K. Drawn initial conditions are afterwards run through the same procedure as described before.

At this point, the second and third step in the fitting procedure can be repeatedly run several times to further pinpoint the best solution. We found out that further refinement was not needed in our case as it did not influence the determined number of stars in the system.

\subsection{Number of stellar components - final classification}
\label{sec:number_stars_tripple}
The fitting procedure described above was used to evaluate observations of every multiple stellar candidate to determine whether they belong to a single, binary or triple stellar system. This resulted in the following set of results for every configuration: predicted \Teff\ of the components, \Feh\ of the system, simulated spectrum, and simulated photometric signature of the system.

As the photometric and spectroscopic fits do not always agree on the best configuration, the following set of steps and rules was applied to classify results in one of six classes presented in Table \ref{tab:res_multiples}.

\begin{table}
	\centering
	\caption{Number of different systems discovered by the fitting procedure performed on possible multiple stars that exhibit excess luminosity.}
	\begin{tabular}{c c}
		\hline
		Configuration classification & Number of systems \\ 
		\hline
		1 star & 2 \\
		$\geq$ 1 star & 14 \\
		2 stars & 27 \\
		$\geq$ 2 stars & 14 \\
		3 stars & 6 \\
		Inconclusive & 1 \\
		\hline
		Total objects & 64 \\
		\hline
	\end{tabular}
	\label{tab:res_multiples}
\end{table}

\begin{itemize}
	\item Compute $\chi^2$ between the simulated photometric signature of the modelled system and extinction corrected absolute photometric observations for every considered stellar configuration. 
	\item Compute $\chi^2$ between the simulated spectrum of the modelled system and the complete GALAH observed spectrum for every considered stellar configuration. 
	\item Independently select the best fitting configuration with the lowest $\chi^2$ for photometric and spectroscopic fit.
	\item If the best photometric and spectroscopic fit point to the same configuration, then the system is classified as having a number of stars defined by both fitting procedures.
	\item If the best photometric and spectroscopic fit do not point to the same configuration, then the system is classified as having at least as many stars as determined by the prediction with a lower number of stars (e.g. $\geq$~2 stars).
	\item If the difference between those two predictions is greater than 1 (e.g. photometric fit points to a single star and spectroscopic to a triple star), then the system is classified as inconclusive. 
\end{itemize}

\begin{figure}
	\centering
	\includegraphics[width=\columnwidth]{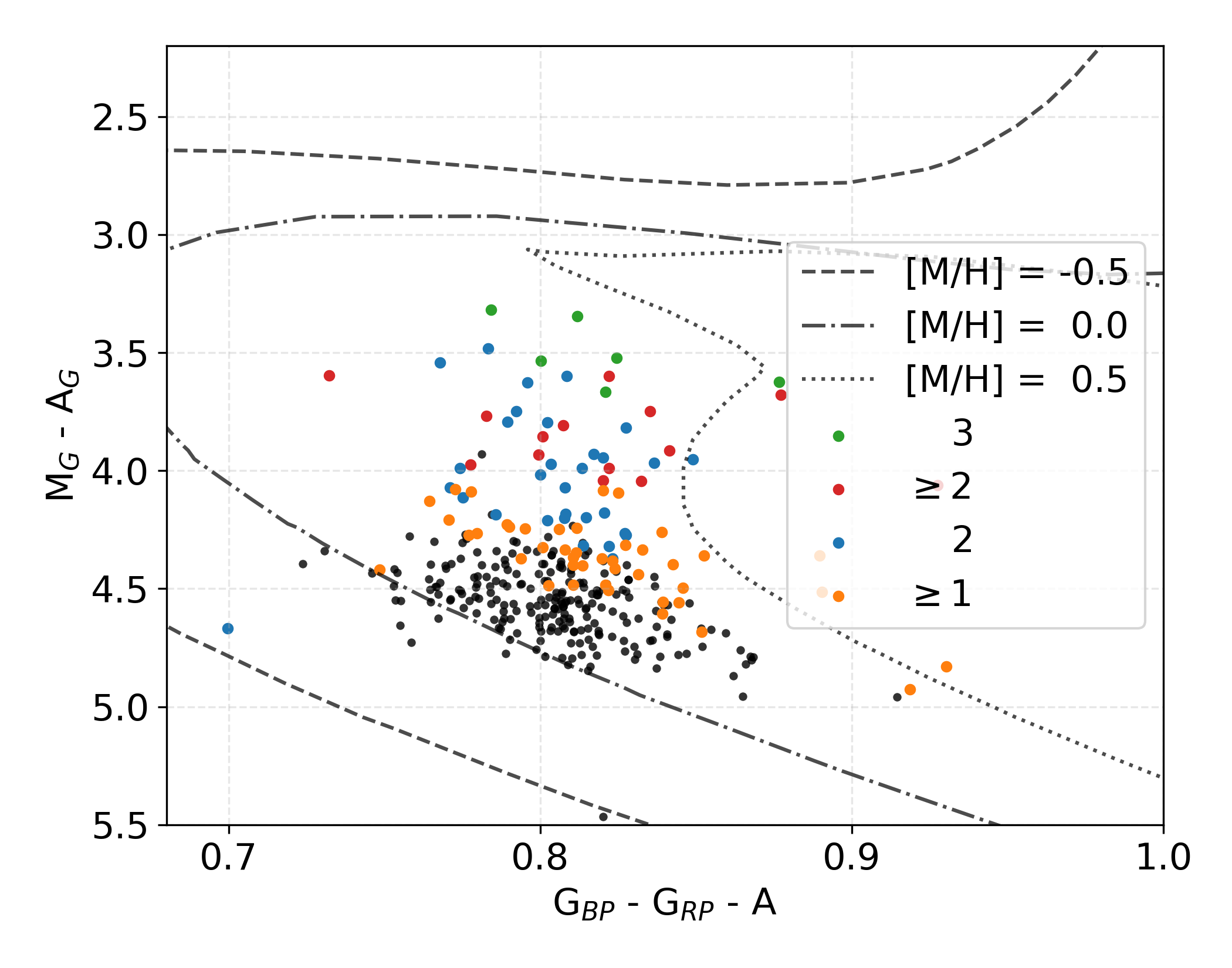}
	\caption{Showing the same data points as Figure \ref{fig:gabs_colour} with indicated definite triple (green dots), possible triple (red dots), binary (blue dots), and possible binary stellar systems (orange dots). All other classes are shown with black dots. Overplotted are PARSEC isochrones \citep{2017ApJ...835...77M} for stars with the age of $4.5$~Gyr and different metallicities.}
	\label{fig:gabs_binmulti}
\end{figure}

The classification produced using these rules is shown as colour coded \G\ colour-magnitude diagram in Figure \ref{fig:gabs_binmulti}.

\subsection{Quality flags}
\label{sec:quaflags}
In additional to our final classification, we also provide an additional quality checks that might help to identify cases for which our method might return questionable determination of a stellar configuration. Every of those checks, listed in Table \ref{tab:q_flag}, is represented by one bit of a parameter \texttt{flag} in the final published table (Table \ref{tab:out_table}). The first bit gives us an indication of whether the object could have an uncertain astrometric solution, whereas the second and the third bits indicate if the final fitted solution has a worse match with the observations than the parameters produced by the \TC\ pipeline. To evaluate this, we used the original stellar parameters reported for the object to construct their photometric and spectroscopic synthetic model that was compared to the observations by computing their $\chi^2$ similarity (\texttt{m\_sim\_p} and \texttt{m\_sim\_f} in Table \ref{tab:out_table}). The resulting fitted spectrum or photometric signature is marked as deviating if its similarity towards observations is worse than for the reported one star parameters. This might not be the best indication of possible mismatch as it is common that \TC\ parameters of the analysed multiple candidates deviate from the main sequence in Kiel diagram (Figure \ref{fig:kiel_cannon}) and therefore fall into less populated parameter space, where they can skew the single star models (Sections \ref{sec:s_model} and \ref{sec:p_model}).

\begin{table}
	\centering
	\caption{Explanation of the used binary quality flags in the final classification of the stellar configuration. A raised bit could indicate possible problems or mismatches in the determined configuration. Symbol X in the last two descriptions represents the best fitting configuration, therefore X~=~[1,2,3].}
	\begin{tabular}{c l}
		\hline
		Raised bit & Description \\ 
		\hline
		0 & None of the flags was raised \\
		1$^{\text{st}}$ bit & High astrometric uncertainty (RUWE~>~1.4)\\
		2$^{\text{nd}}$ bit & Deviating photometric fit (\texttt{sX\_sim\_p} > \texttt{m\_sim\_p})\\
		3$^{\text{rd}}$ bit & Deviating spectroscopic fit (\texttt{sX\_sim\_s} > \texttt{m\_sim\_s})\\
		\hline
	\end{tabular}
	\label{tab:q_flag}
\end{table}

\begin{figure}
	\centering
	\includegraphics[width=\columnwidth]{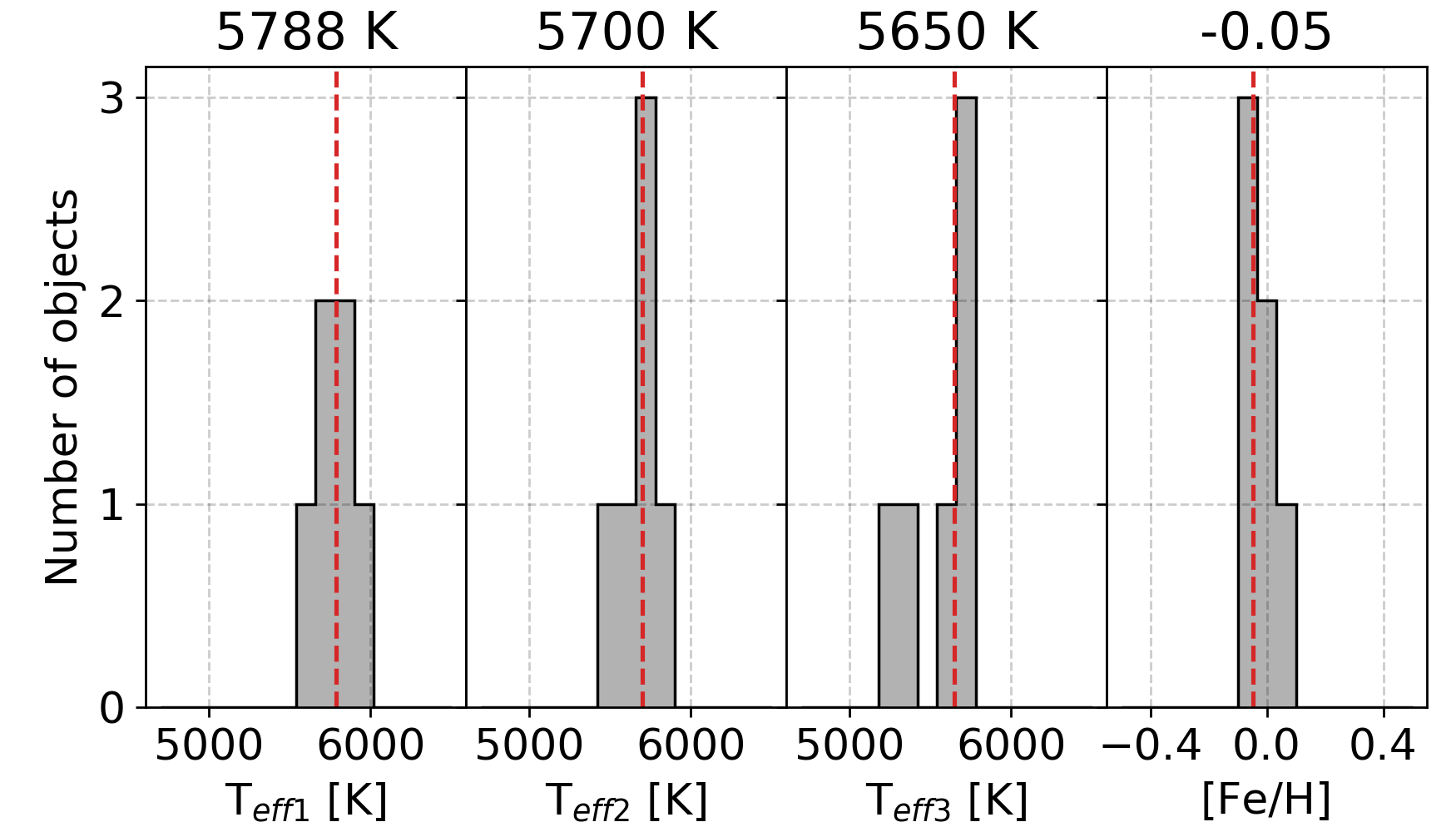}
	\caption{Parameters of triple stellar systems discovered and characterized by our analysis. Histograms represent distribution of all fitted results for 6 objects that were classified as triple stellar systems and observationally mimic Solar spectrum. Median values of the distributions are given above individual histogram and indicated with dashed vertical line.}
	\label{fig:triple_params}
\end{figure}

\section{Characterization of single star candidates}
\label{sec:single_fit}
Once we concluded our analysis of the set of 64 objects that showed obvious signs of excess luminosity, we then proceeded to study the remaining 265 objects that are most probably not part of a complex stellar system. To explore their composition, they were analysed with the same procedure as multiple candidates (Section \ref{sec:multi_fit}). Before running the procedure, we omitted the option to fit for a triple system as they clearly do not possess enough excess luminosity for that kind of a system.

The obtained results were analysed and classified using the same set of rules introduced in Section \ref{sec:number_stars_tripple}. Retrieved classes are summarized in Table \ref{tab:res_single} and in Figure \ref{fig:gabs_binmulti}. In the latter we see that the potential binaries are located on the top of the colour-magnitude diagram, which is consistent with the potential presence of an additional stellar source.

\begin{table}
	\centering
	\caption{Number of different systems discovered by the fitting procedure performed on stars that do not exhibit excess luminosity. Classes and their description is the same as used in Table \ref{tab:res_multiples}. In addition, number of stars without parallactic measurements is added for completeness.}
	\begin{tabular}{c c}
		\hline
		Configuration classification & Number of systems \\ 
		\hline
		1 star & 230 \\
		$\geq$ 1 star & 31 \\
		2 stars & 4 \\
		$\geq$ 2 stars & 0 \\
		3 stars & 0 \\
		Inconclusive & 0 \\
		Unknown parallax & 0 \\
		\hline
		Total objects & 265 \\
		\hline
	\end{tabular}
	\label{tab:res_single}
\end{table}

\section{Orbital period constraints}
\label{sec:orital_periods}
The observational data we have gathered on possible multiple systems point to configurations that change slowly as a function of time. Using the observational constraints and models that describe the formation of the observed spectra, we try to set limiting values on the orbital parameters of the detected triple stellar system candidates. In order to have a greater sample size, we use both definite (class 3) and probable (class $\geq$2) triple stars.

With the limited set of observations, we have to set assumptions about the constitution of those systems. For a hierarchical 2~+~1 system to be dynamically stable on long timescales, its ratio between the orbital period of an inner pair $P_S$ and outer pair $P_L$ must be above a certain limit. \citet{2006epbm.book.....E} showed that $P_L$/$P_S$ must be higher than 5. The same lower limit is noticeable when the correlation between periods of the known triple stars is plotted \citep{2008MNRAS.389..925T, 2018ApJS..235....6T}.

In order to estimate the periods, we have to know the masses and distances between the stars in a system. Without complete information about the projected velocity variation in a system, masses can also be inferred from the spectral type. As we are looking at the Solar twin triples, whose effective temperatures are all very similar and close to Solar values (see Figure \ref{fig:triple_params}), our rough estimate is that all stars also have a Solar-like mass $\sim$M$_{\sun}$. From this, we can set the inner mass ratio $q_S$ to be close to $1$ and the outer mass ratio $q_L \sim 0.5$. When we are dealing with the outermost star, the inner pair is combined into one object with twice the mass of the Sun. The likelihood of such a configuration is also supported by the observations \citep{2008MNRAS.389..925T} where a higher concentration of triple systems is present around those mass ratios. Contrary to our systems, twin binaries with equal masses usually have shorter orbital periods \citep{2000A&A...360..997T}. 

With the initial assumption about the configuration of the triple systems and masses of the stars, the periods of the inner and outer pair can be constrained to some degree as other orbital elements (inclination, ellipticity, phase \ldots) are not known.

\begin{figure}
	\centering
	\includegraphics[width=\columnwidth]{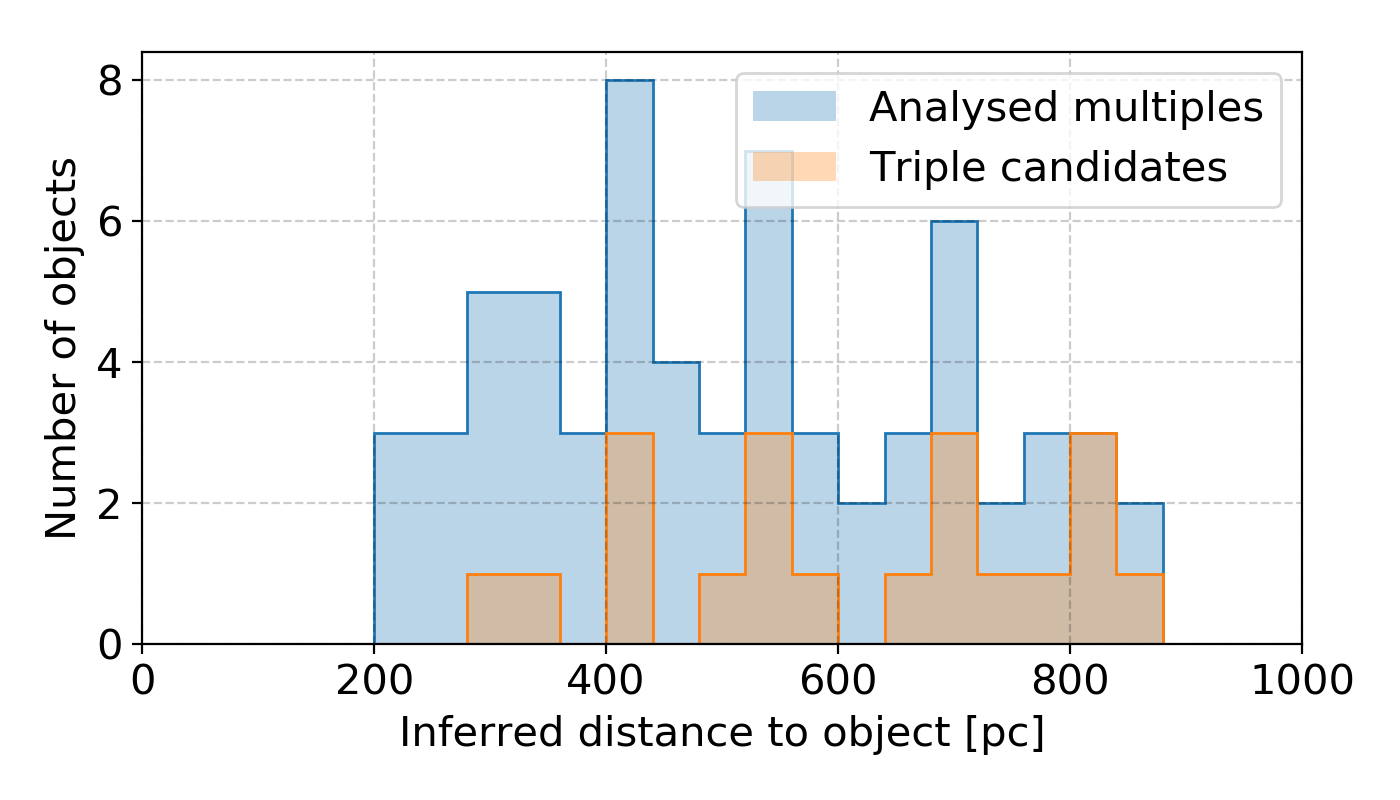}
	\caption{Histogram showing distribution of radial distances for the analysed set of stars. Distances were inferred by the Bayesian approach that takes into account the distribution of stars in the Galaxy \citep{2018AJ....156...58B}.}
	\label{fig:gaia_dist}
\end{figure}

\subsection{Outer pair and \G\ angular resolution}
\label{sec:orbits_gaia}
Limited by its design, \G\ spacecraft and its on-ground data processing are in theory unable to resolve stars with the angular separation below $\sim0.1$~arcsec. Above that limit, their separability is governed by the flux ratio of the pair. The validation report of the latest data release \citep{2018A&A...616A..17A} shows that the currently achieved angular resolution is approximately $0.4$~arcsec as none of the source pairs is found closer than this separation. We used this reported angular limit to assess possible orbital configurations that are consistent with both the spectroscopic and astrometric observations.

Along with the final astrometric parameters, the \G\ data set also contains information about the goodness of the astrometric fit. \citet{2018RNAAS...2b..20E} used those parameters to confirm old and find new candidates for unresolved exoplanet hosting binaries in the data set. As we are looking at a system of multiple stars that orbit around their common centre of mass, we expect their photo-centre to slightly shift during such orbital motion. The observed wobble of the photo-centre also depends on the mass of stars in the observed system. In the case of a binary system containing two identical stars, the wobble would not be observed, but its prominence increase as the difference between their luminosities becomes more pronounced. This subtle change in the position of a photo-centre adds additional stellar movement to the astrometric fit, consequently degrading its quality. Improvements for such a motion will be added in latter \G\ data releases. If a system has an orbital period much longer than the time-span of \G\ DR2 measurements (22 months), its movement has not yet affected the astrometric solution. This puts an upper limit on an orbital period as it should not be longer than few years in order to already affect the astrometric fit results. 

Setting limits to the parameters of astrometric fit quality \citep[\texttt{astrometric\_excess\_noise}~>~5, and \texttt{astrometric\_gof\_al}~>~20 as proposed by][]{2018RNAAS...2b..20E}, none of our 329 stars meets those requirements. This suggests that all of them are most likely well below the \G\ separability limit and/or have long orbital periods. Another indicator for a lower-quality astrometric fit that we can use is RUWE. Figure \ref{fig:gabs_colour} shows distribution of potentially problematic large RUWE among single and multiple candidates. The latter, on average, have a much poorer fit quality that might indicate a presence of an additional parameter that needs to be considered in the astrometric fit. 

Distances to triple stars, shown by the histogram in Figure \ref{fig:gaia_dist} range from around $0.3$ to $0.9$~kpc. From this we can assume the maximal allowable distance between components of an outer pair to be in the order of $100-350$~AU, pointing to outer orbital periods larger than $500$~years. To test if such systems would meet our detection constraints, we created 100,000 synthetic binary systems whose orbital parameters were uniformly distributed within the parameter ranges given in Table \ref{tab:mc_gaia}. Observable radial velocities of both components were computed using the following equations:

\begin{equation}
	\label{equ:rad_vel1}
	v_{1} = \frac{2 \pi \sin i}{P \sqrt{1 - e^2}} \frac{a q}{1+q} \Big(\cos(\theta + \omega) + e \cos \omega \Big)
\end{equation}

\begin{equation}
	\label{equ:rad_vel2}
	v_{2} = \frac{2 \pi a \sin i}{P \sqrt{1 - e^2}} \frac{a}{1+q} \Big(\cos(\theta + \omega + \pi) + e \cos (\omega  + \pi) \Big)
\end{equation}

\begin{equation}
	\label{equ:rad_period}
	P = \sqrt{a^3 \frac{4 \pi^2}{G M_1 (1 + q)}}
\end{equation}

\begin{figure}
	\centering
	\includegraphics[width=\columnwidth]{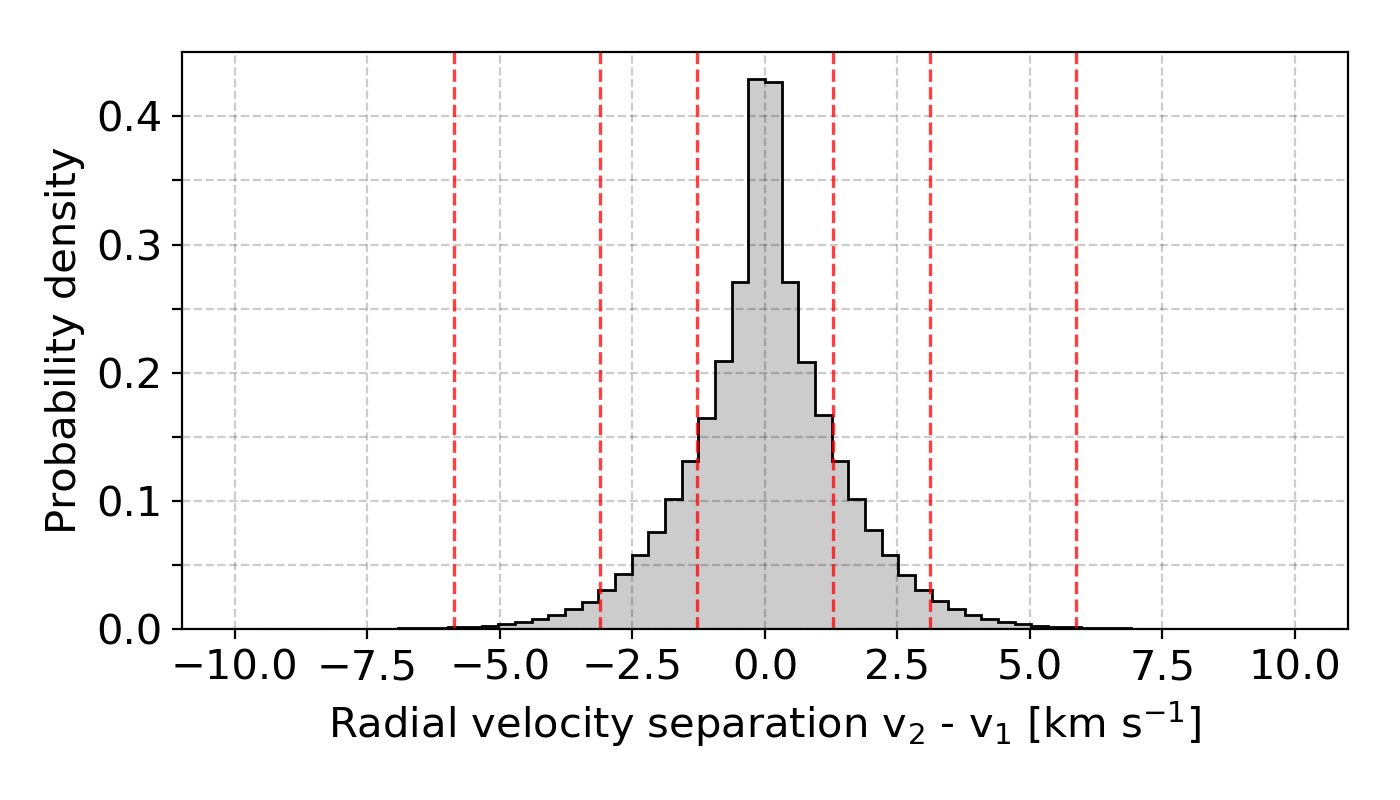}
	\caption{Distribution of computed radial velocity separations between a primary and secondary component of the simulated binary systems defined by the orbital parameters given in Table \ref{tab:mc_gaia}. Red vertical lines show 1, 2, and 3 $\sigma$ probabilities of the given distribution.}
	\label{fig:mc_rv_sep}
\end{figure}

The distribution of velocity separations ($\Delta v = v_2 - v_1$) for synthetic systems is given by Figure \ref{fig:mc_rv_sep}, where we can see that more than $99.7$~\% of generated configurations would produce a spectrum that would still be considered as a Solar twin ($\Delta v$~<~6~\kms, see Section \ref{sec:rv_sep_sim} and Figure \ref{fig:rv_similarity_2} for further clarification). If we set the semi-major axis $a$ (in Table \ref{tab:mc_gaia}) to single value in the same simulation, we can find the closest separation that would still meet the same observational criteria in at least $68$~\% of the cases. In our simulation this happens at the mutual separation of $10$~AU (and at $50$~AU for $95$~\% of the cases). The orbital period of a such outer pair is about $18$~years (and $200$~years for $50$~AU) and is most probably way too long to had significant effect on the quality of the astrometric fit. Considering observationally favorable orbital configurations with face-on orbits, the actual system could be much more compact than estimated.

\begin{table}
	\centering
	\caption{Ranges of the orbital parameters used for the prediction of observable radial velocity separation between stars in an outer binary pair. The range of the semi-major axis length $a$ is set between the \G\ separability limit for the closest and farthest triple candidate. The uniform distribution of $a$ is a good approximation of the real periodicity distribution published by \citet{2010ApJS..190....1R} as we are sampling a narrow range of it. Use of the real observed distribution would in our case introduce insignificant changes in radial velocity separation as we are simulating wide, slowly rotating systems.}
	\begin{tabular}{c | c}
		\hline
		Parameter & Considered range \\ 
		\hline
		$M_1$ & 2 M$_{\sun}$ \\
		$a$ & 100 \ldots 350 AU \\
		$q$ & 0.45 \ldots 0.55 \\
		$\sin i$ & 0 \ldots 1 (i = 0 \ldots 90 deg)\\
		$e$ & 0.1 \ldots 0.8 \\
		phase & 0 \ldots 1, used for calculation of $\theta$ \\
		$\omega$ & 0 \ldots 360 deg \\
		\hline
	\end{tabular}
	\label{tab:mc_gaia}
\end{table}

\subsection{Inner binary pair and formation of double lines in a spectrum}
\label{sec:orbits_sb2}
\begin{figure}
	\centering
	\includegraphics[width=\columnwidth]{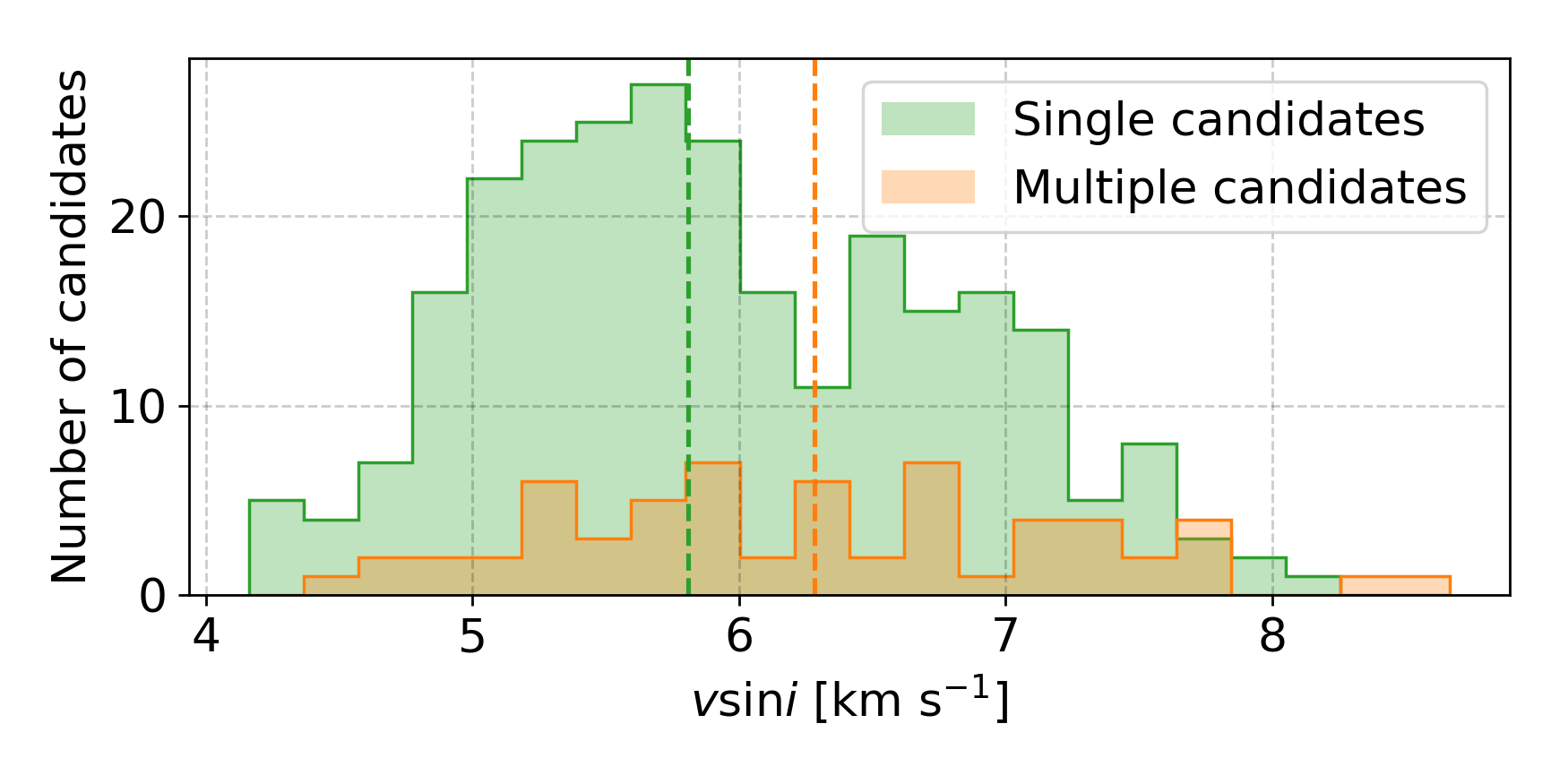}
	\caption{Distribution of determined \vsin\ for analysed single and multiple candidates. Median values of the distributions are marked with dashed vertical lines.}
	\label{fig:vsini_hist}
\end{figure}

An upper estimate of orbital sizes for an outer pair gives us a confirmation that almost every considered random orbital configurations would satisfy the observational and selection constraints. To ensure the long-term stability of such a system, the inner pair must have a period that is at least five times shorter \citep{2006epbm.book.....E}. At such short orbital periods, the inner stars could potentially move sufficiently rapidly in their orbits as to produce noticeable absorption line splitting for edge-on orbits. To be confident that none of the analysed objects produces such an SB2 spectrum, we visually checked all considered spectra and found no noticeable line splitting in any of the acquired GALAH or Asiago spectra. A subtle hint about a possible broadening of the spectral lines comes from the determined \vsin. The median of its distribution in Figure \ref{fig:vsini_hist} is higher for multiple candidates with excess projected velocity of $\sim$0.5~\kms.

Accounting for the GALAH resolving power and spectral sampling, we can estimate a minimal radial velocity separation between components of a spectrum to show clear visual signs of duplicated spectral lines. To determine a lower limit, we combined two Solar spectra of different flux ratios and visually evaluated when the line splitting becomes easily noticeable. With equally bright sources, this happens at the separation of $\sim$14~\kms. When the secondary component contributes 1/3 of the total flux, minimal separation is increased to $\sim$20~\kms. A similar separation is needed when secondary contributes only 10~\% of the flux. As no line splitting is observed in our analysed spectra, we can be confident that the velocity separation between the binary components was lower than that during the acquisition.

\begin{figure}
	\centering
	\includegraphics[width=\columnwidth]{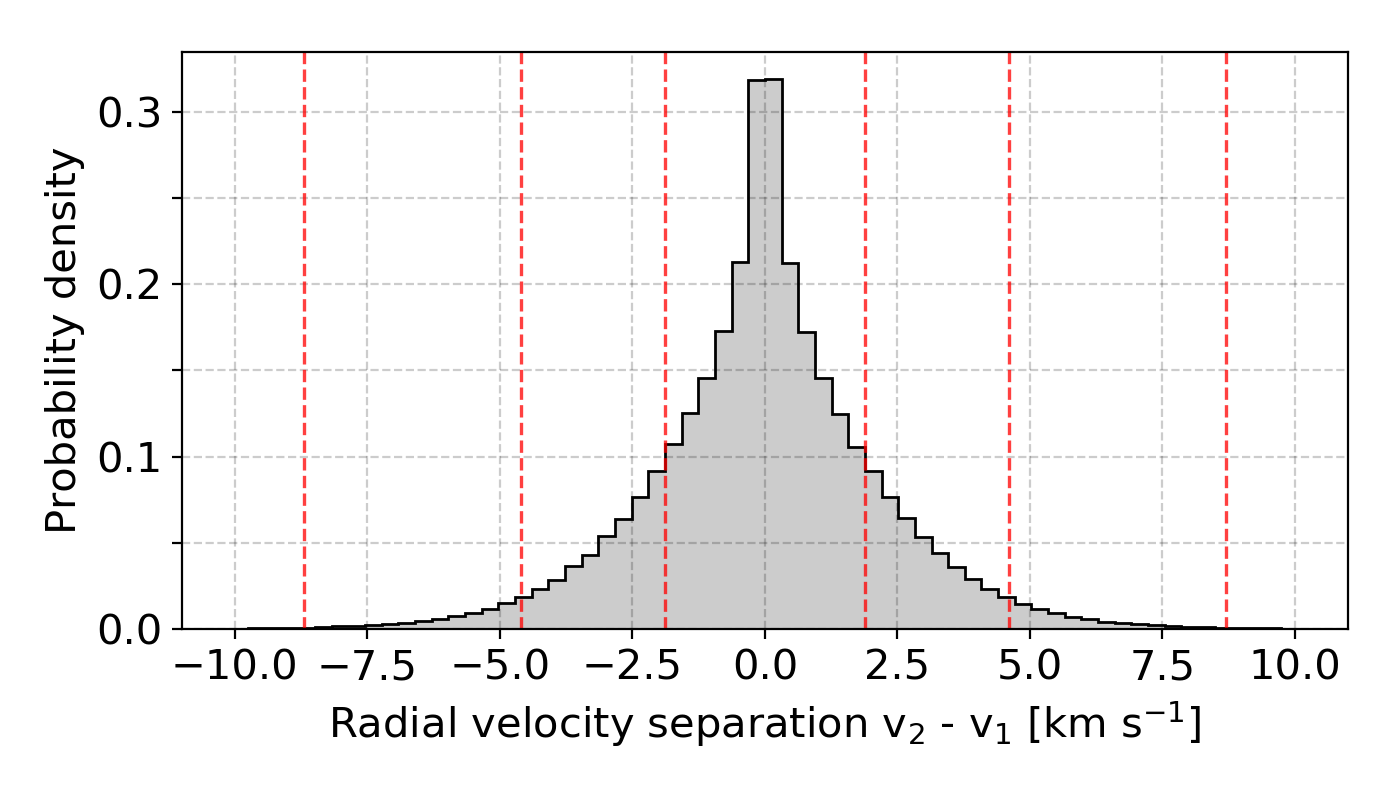}
	\caption{Same as Figure \ref{fig:mc_rv_sep}, but for an inner binary of a simulated triple stellar system.}
	\label{fig:mc_rv_sep_inner}
\end{figure}

Considering the minimal ratio between outer and inner binary period, we can deduce an expected radial velocity separation of an inner pair for the widest possible orbits. As explained in previous section, we used Equations \ref{equ:rad_vel1}-\ref{equ:rad_period} and possible ranges of inner orbital parameters (Table \ref{tab:mc_gaia_inner}) to generate a set of synthetic binary systems. The distribution of their $\Delta v$ is shown in Figure \ref{fig:mc_rv_sep_inner} and represent inner binaries with orbital periods from $100$ to $700$~years. At those orbital periods, more that $92$~\% of the considered configurations would satisfy the condition of $\Delta v$~<~4~\kms, ensuring that the observed composite of two equal Solar spectra would still be considered as a Solar twin (see Section \ref{sec:rv_sep_sim} and Figure \ref{fig:rv_similarity} for further clarification). If we limit our synthetic inner binaries to only one orbital period, we can estimate the most compact system that would still meet the observational criteria in at least $68$~\% of the cases. This would happen at the orbital period of $40$~years with a semi-major axis of $14$~AU. 

\begin{table}
	\centering
	\caption{Same as Table \ref{tab:mc_gaia}, but for an inner binary of a hierarchical triple stellar system.}
	\begin{tabular}{c | c}
		\hline
		Parameter & Considered range \\ 
		\hline
		$M_1$ & M$_{\sun}$ \\
		$P_S$ & $P_L$ / 5, used for calculation of $a$ \\
		$q$ & 0.9 \ldots 1.0 \\
		$\sin i$ & 0 \ldots 1 (i = 0 \ldots 90 deg)\\
		$e$ & 0.1 \ldots 0.8 \\
		phase & 0 \ldots 1, used for calculation of $\theta$ \\
		$\omega$ & 0 \ldots 360 deg \\
		\hline
	\end{tabular}
	\label{tab:mc_gaia_inner}
\end{table}

\subsection{Multi-epoch radial velocities}
\label{sec:orbits_rv}
\begin{figure}
	\centering
	\includegraphics[width=\columnwidth]{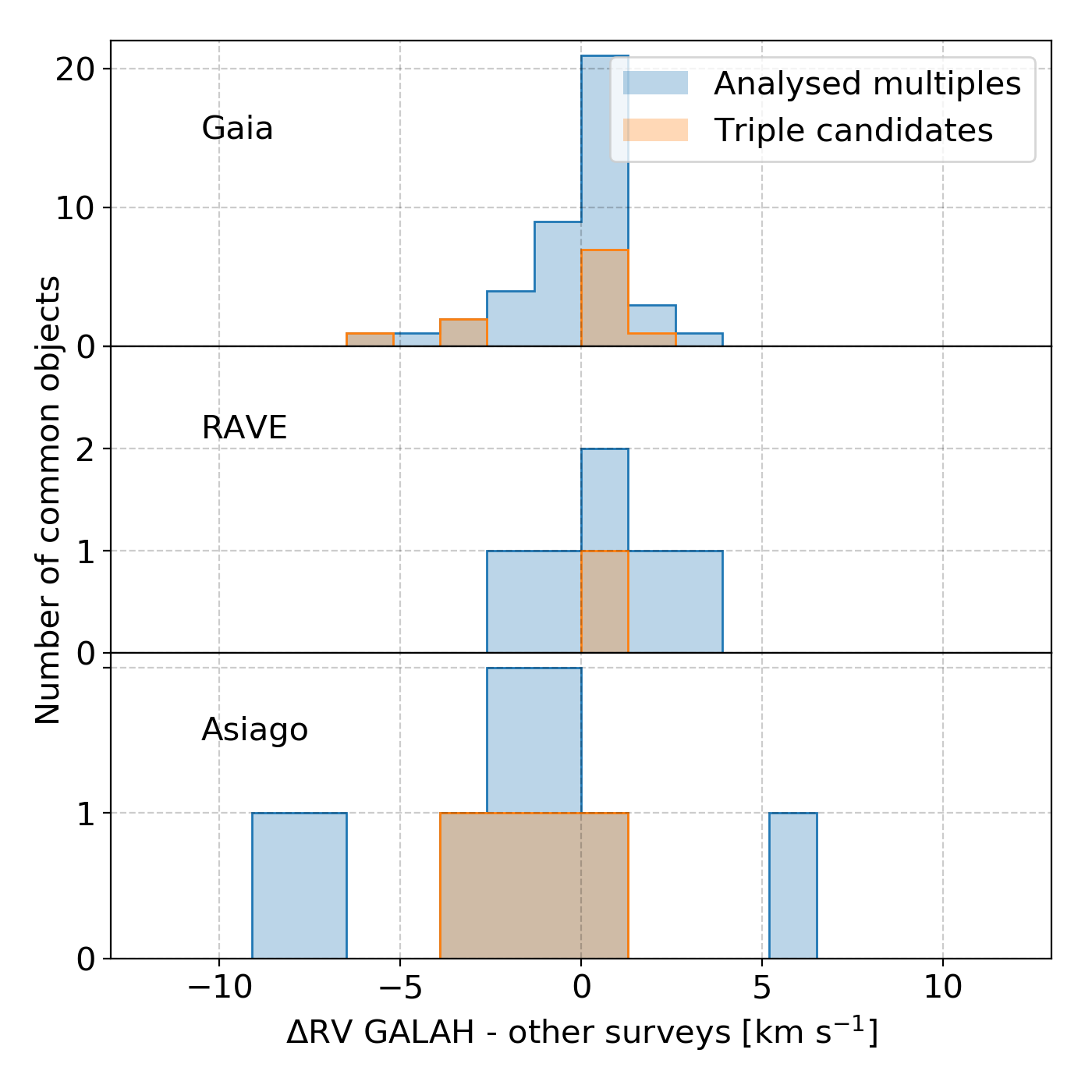}
	\caption{RV difference between GALAH and other large spectroscopic surveys. The name of an individual survey or observatory is given in the upper-left corner of every panel. Blue histograms represent velocity difference for all investigated multiple candidates and orange histograms for discovered triple systems only.}
	\label{fig:rv_survey}
\end{figure}

To support our claims about slowly changing low orbital speeds in detected triple candidates, we analysed changes in their measured radial velocities between the GALAH and other comparable all-sky surveys. Distributions of changes are presented by three histograms in Figure \ref{fig:rv_survey} where almost all velocity changes, except one, are within 5~\kms. Differences between the GALAH and \G\ radial velocities were expected to be small as the latter reports median velocities in the time-frame that is similar to the acquisition span of the GALAH spectra. Extending the GALAH observations in the past with RAVE and in the future with Asiago observations did not produce any extreme changes. For this comparison, we also have to consider the uncertainty of the measurements that are in the order of $\sim$2~\kms\ for RAVE and $\sim$1~\kms\ for Asiago spectra.

With the data synergies, we could produce only three observational time-series that have observations at more than two sufficiently separated times. With only three data points in each graph, not much can be said about actual orbits.

\section{Simulations and tests}
\label{sec:simulations}
To determine the limitations of the employed algorithms, we used them to evaluate a set of synthetic photometric and spectroscopic sources. As the complete procedure depends on the criteria for the selection of Solar twin candidates, we first investigated if the selection criteria allows for any broadening of the spectral lines or their multiplicity as they are both signs of a multiple system with components at different projected radial velocities.

In the second part, we generated a set of ideal synthetic systems that were analysed by the same fitting procedure as the observed data set. The results of this analysis are used to determine what kind of systems could be recognized with the fitting procedure and how the results could be used to spot suspicious combinations of fitted parameters. 

In the final part, we try to evaluate selection biases that might arise from the position of analysed stars on the \G\ colour-magnitude diagram and from the GALAH selection function that picks objects based on their apparent magnitude.

\subsection{Radial velocity separation between components}
\label{sec:rv_sep_sim}
To determine the minimal detectable radial velocity (RV) separation between components in a binary or a triple system, we constructed a synthetic spectrum resembling an observation of multiple Suns at a selected RV separation. The spectrum of a primary component was fixed at the rest wavelength with RV~=~0~\kms\ and a secondary spectrum shifted to a selected velocity. After the shift, these two spectra were added together based on their assumed flux ratio. 

The generated synthetic spectrum was compared to the Solar spectrum with exactly the same metric as described in Section \ref{sec:solar_twins_sel}. Computed spectral similarities at different separations were compared to the similarities of analyzed Solar twin candidates. The first RV separation that produces a spectrum that is more degraded than the majority of Solar twin candidates was determined to be a minimal RV at which the observed spectrum would be degraded enough that it would no longer be recognized as a Solar twin. The high SNR of the generated spectrum was not taken into account for this analysis in contrast to the algorithm that was used to pinpoint Solar twin candidates. Therefore we also omitted candidates with a lower similarity that in our case directly corresponds to their low SNR. 

\begin{figure}
	\centering
	\includegraphics[width=\columnwidth]{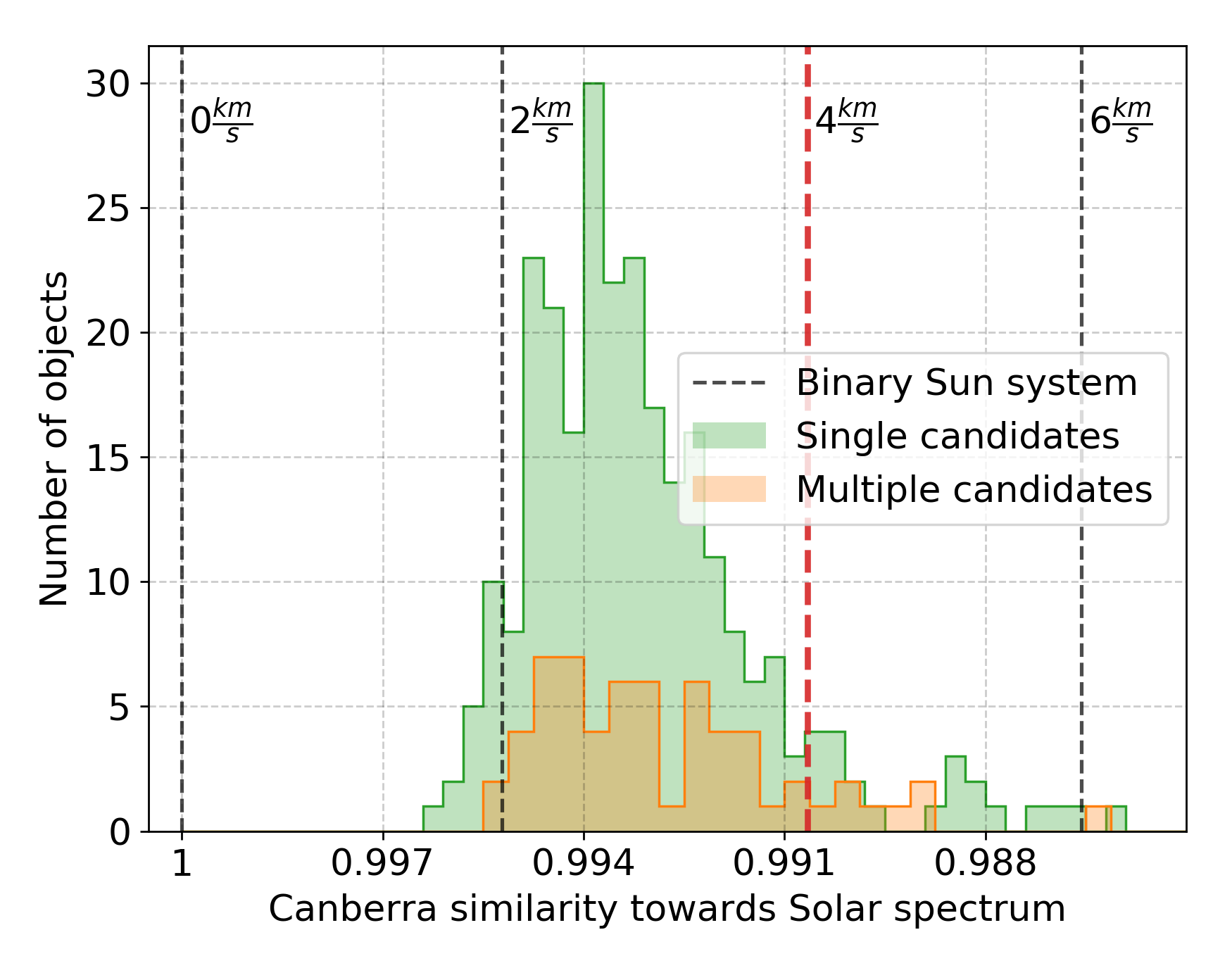}
	\caption{Histograms of Canberra similarities towards Solar spectrum for multiple stellar candidates (orange histogram) and single star candidates (green histogram) in the red HERMES arm. Vertical dashed lines represent the same similarity measure, but for a binary system comprising of two equally bright Suns, whose components are at different radial velocities. The separation between components is increasing in 2~\kms\ steps, where the leftmost vertical line represents the case where both stars are moving with the same projected velocity. The selected maximal velocity at which the composite spectrum would still be considered a Solar twin is marked with the thicker red vertical line. Distribution of histograms also shows that spectra of multiple system candidates are as (dis)similar as spectra of single stars.}
	\label{fig:rv_similarity}
\end{figure}

The result of this comparison is presented in Figure \ref{fig:rv_similarity}, where we can see that the minimal detectable separation of two equally bright stars resembling the Sun is $\sim$4~\kms. In the case where a primary star contributes $2/3$ of the total flux, the minimal RV increases to $\sim$6~\kms\ (see Figure \ref{fig:rv_similarity_2}). Further increase in the ratio between their fluxes would also increase a minimal detectable separation, but only to a certain threshold from where on a secondary star would not contribute enough flux for it to be detectable, and its received flux would be comparable to the typical HERMES spectral noise. In our case, this happens when the secondary contributes less than 10~\% of the total flux. These boundaries are only indicative as they also heavily depend on the quality of the acquired spectra. When a low level of noise with the Gaussian distribution ($\sigma=0.01$) is added to a secondary component with a comparable luminosity, the minimal RV decreases because the similarity between spectra also decreases. In that case, the similarity for $\Delta v$~=~0 is located near the mode value of similarity distribution in Figure \ref{fig:rv_similarity}. 

\subsection{Analysis of synthetic multiple systems}
\begin{figure}
	\centering
	\includegraphics[width=\columnwidth]{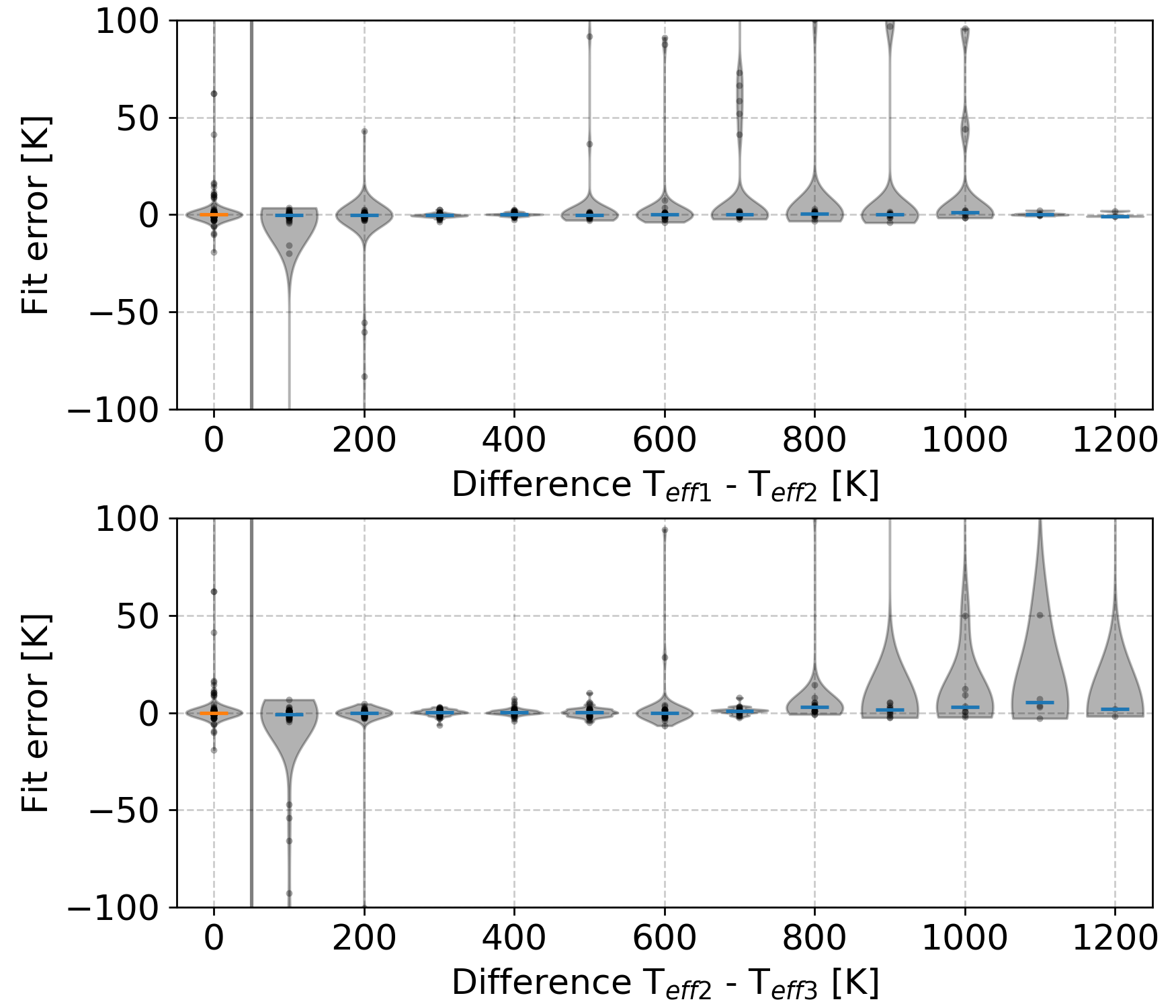}
	\caption{The accuracy of our analysis when the fitting procedure is applied to a synthetic triple system. Upper panel shows the distribution of \Teff\ prediction errors for a secondary star depending on the difference between selected temperatures of a primary and a secondary star. The lower panel shows the same relation but for a tertiary star in comparison to a secondary. As a reference, the prediction error of a primary star is shown on the left side of both panels. Labels \Teffn{1}, \Teffn{2}, and \Teffn{3} indicate decreasing effective temperatures of stars in a simulated system.}
	\label{fig:triple_sym_res}
\end{figure}

The limitations and borderline cases of the fitting procedure were tested by evaluating its performance on a set of synthetic binary and triple systems that were generated using the same models and equations as described in the fitting procedure. An ideal, virtually noiseless set comprised 95 binary and 445 triple systems whose components differ in temperature steps of 100~K. The hottest component in the system was set to values in the range between 5300 and 6200~K. The \Teff\ of the coldest component could go as low as 4800~K. The \Feh\ of all synthetic systems was set to $0.0$ to mimic Solar-like conditions. Condensed results are presented in Figures \ref{fig:triple_sym_res} and \ref{fig:binary_sym_res}.

As expected, the fitting procedure (Section \ref{sec:multi_fit}) did not have any problem determining the correct configuration of the synthesized system. Temperatures of the components were also correctly recovered with a median error of $0 \pm 13$~K for the hottest component and  $0 \pm 43$~K for the coldest component in a triple system. A more detailed analysis with the distribution of prediction errors, where the temperature difference between components is taken into account, is presented in Figure \ref{fig:triple_sym_res}. From that analysis, we can deduce that \Teff\ of the secondary component is successfully retrieved if it does not deviate by more than 1000~K from the primary. Results at such large temperature difference are inconclusive as the number of simulated systems drops rapidly. The same can be said for the tertiary component, but with the limitation that it should not be colder by more than 700~K when compared to the secondary star. Beyond that point, the uncertainty of the fitted result increases and a star is determined to be hotter as it really is.

Another application of such an analysis is to identify signs that could point to a possibly faulty solution when it is comparably likely that a multiple system is comprised of two or three components. The results of such an analysis are presented in Figure \ref{fig:binary_sym_res}, where we tried to describe a binary system with a triple system fit. From the distribution of errors, we can observe that the effective temperature of a primary star with the largest flux is recovered with the smallest fit errors whose median is $20$~$\pm$~$93$~K. As we are using too many components in that fit, \Teff\ of a secondary is reduced in order to account for the redundant tertiary component in the fit. As imposed by the limit in the fitting procedure, the tertiary \Teffn{3} is set to as low as possible. If the same thing happens for a real observed system, this could be interpreted as a model over-fitting.

\subsection{Triple stars across the H-R diagram}
In the current stage of Galactic evolution, binary stars with a Solar-like \Teff\ are located near a region that is also occupied by main sequence turnoff stars in the Kiel and colour-magnitude diagram (Figures \ref{fig:kiel_cannon} and \ref{fig:gabs_binmulti}). This, combined with the fact that older stars with a comparable initial mass and metallicity would also pass a region occupied with binaries (Figure \ref{fig:gabs_colour}) poses an additional challenge for detection of unresolved multiples if their spectrum does not change sufficiently during the evolution.

\begin{figure}
	\centering
	\includegraphics[width=\columnwidth]{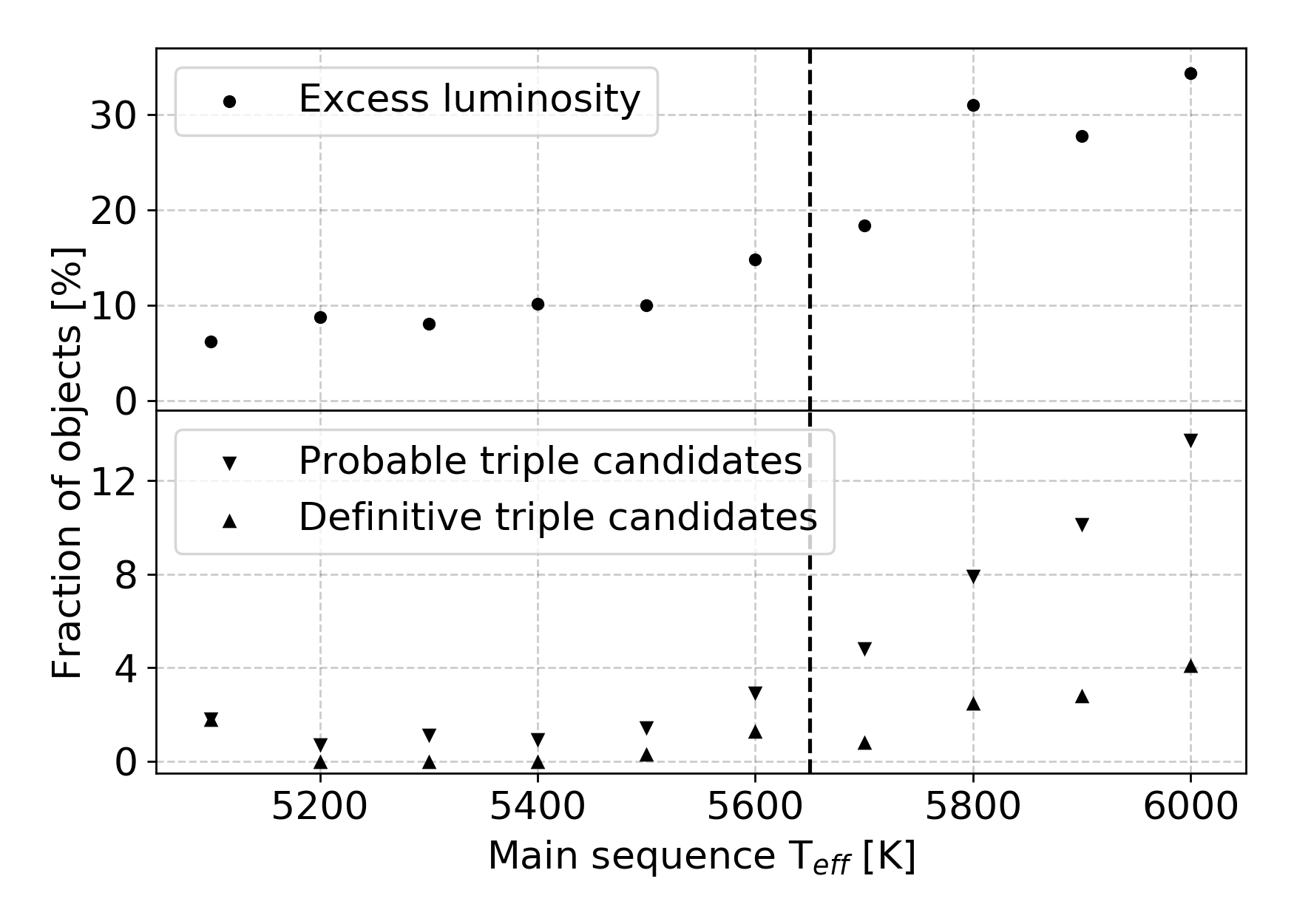}
	\caption{The upper panel shows the percentage of objects at different positions on the Kiel diagram shown in Figure \ref{fig:kiel_cannon} that show excessive luminosity and are spectroscopically similar to main sequence stars. Similarly, the lower panel shows upper and lower boundary on a percentage of triple system candidates at the same positions. For a definitive candidate, both fits must agree on a triple configuration. The strong dashed vertical line repents a point on Kiel diagram where the main sequence visually starts merging with the red-giant branch, the point where a region above the main sequence becomes polluted with more evolved stars.}
	\label{fig:triple_hr}
\end{figure}

To analyse the possible influence of the turnoff stars and older, more evolved stars, we ran the same detection procedure as described in Section \ref{sec:solar_twins_sel} and analysis (Section \ref{sec:multi_fit}) to determine the fraction of binaries and triples at different \Teff, ranging from 5100 to 6000~K, with a step of 100~K. A comparison median spectrum was computed from all spectra in the range $\Delta$\Teff~$\pm$~60~K, $\Delta$\Logg~$\pm$~0.05~dex and $\Delta$\Feh~$\pm $~$0.05$~dex, where the main sequence \Logg\ was taken from the main sequence curve shown in Figure \ref{fig:kiel_cannon}, and \Feh\ was set to $0.0$. The limiting threshold for multiple candidates was automatically determined in the same way as for Solar twin candidates (Section \ref{sec:multi_cand}).

Condensed results of the analysis are shown in Figure \ref{fig:triple_hr}, where we can see that both the fraction of stars with excessive luminosity and triple candidates starts increasing above the \Teff~$\sim$~5600~K. This might indicate that the underlying distribution of more evolved and/or hotter stars might have some effect on our selection function. On the other hand, this increasing binarity fraction coincides with other surveys that show similar trends \citep{2013ARA&A..51..269D}.

As the detected triple star candidates encompass a fairly small region in a colour-magnitude space, we were interested in the degree to which our analysed spectra are similar to those of other stars in the same region. For this purpose, the GALAH objects with absolute \G\ corrected magnitudes in the range $3.3$~<~M$_{G}$~<~$3.6$ and $0.77$~<~G$_{BP}$-G$_{RP}$~<~$0.84$ were selected and plotted in Figure \ref{fig:tsne_marked}, where spectra are arranged according to their similarity. In this 2D projection \citep[details about the construction of which are given in][]{2017ApJS..228...24T, buder2018} it is obvious that spectra considered in this paper clearly exhibit a far greater degree of mutual similarity than other spectra with similar photometric signature. As expected, many of them lie inside the region of SB2 spectra. All of our Solar twin candidates were visually checked for the presence of a resolvable binary component. Nevertheless, this plot gives us additional proof of their absence as none of the analysed twins is located inside that region.

\subsection{Observational bias - Galaxia model}
\label{sec:bias}
Every magnitude limited survey, such as the GALAH, will introduce observational bias into the frequency of observed binary stars as their additional flux changes the volume of the Galaxy from where they are sampled. Their distances and occupied volume of space is located further from the Sun and therefore also greater than for comparable single stars with the same apparent magnitude and colour.

To evaluate this bias for our type of analysed stars, we created a synthetic Milky Way population of single stars using the Galaxia code \citep{2011ApJ...730....3S}. First, the code was run to create a complete population of stars in the apparent V magnitude range between 10 and 16 without any colour cuts. In order for the distribution of synthesized stars to mimic the GALAH survey as closely as possible, only stars located inside the observed GALAH $2^\circ$ fields were retained. After the spatial filtering, only stars with Sun-like absolute magnitude $M_V~=~4.81$~$\pm$~$0.05$ and colour index $B-V$~=~$0.63$~$\pm$~$0.05$ were kept in the set \citep[reference magnitudes were taken from][]{2018ApJS..236...47W}.

From the filtered synthetic data set we took three different sub-sets representing a pool of observable single Solar twins ($12.0$~$\leq$~V~$\leq$~$14.0$), binary twins ($12.75$~$\leq$~V~$\leq$~$14.75$), and triple twins ($13.2$~$\leq$~V~$\leq$~$15.2$) based on their apparent V magnitudes. Extinction and reddening were not used in this selection as their use did not significantly alter the relationship between the number of stars in sub-sets. Assuming that the frequency of multiple stars is constant and does not change in any of those sets, we can estimate the selection bias on the frequency of multiple stars. The number of stars in each sub-set was 15862, 35470, and 54567 respectively. According to those star counts, the derived frequency of binaries would be too high by a factor of $2.2$ and a factor $3.4$ for triple stars. Considering those factors, the fraction of unresolved triple candidates with Solar-like spectra is $\sim$2~\% and $\sim$11~\% for binary candidates. 

Accidental visual binaries that lie along the same line-of-sight and have angular separation smaller that the field-of-view of the 2dF spectroscopic fiber ($2 \arcsec$) or smaller than the \G\ end-of-mission angular resolution ($0.1 \arcsec$) were not considered in this estimation.

\section{Conclusions}
\label{sec:concl}
Combining multiple photometric systems with spectroscopic data from the GALAH and astrometric measurements taken by \G, we showed the possible existence of triple stellar systems with long orbital periods whose combined spectrum mimics Solar spectrum. The average composition of such a system consists of three almost identical stars, where one of the stars is $\sim$10~K warmer than the Sun and the coldest has an effective temperature $\sim$120~K below the Sun. The derived percentage of such unresolved systems would be different for nearby/close stars as they become more/less spatially resolved. In the scope of our magnitude limited survey, we sampled only a fraction of possible distances to the systems.

Without any obvious signs of the orbital periodicity in the measured radial velocities of the systems, orbital periods were loosely constrained based on the observational limits and few assumptions. By the prior assumption that the \G\ spacecraft sees those systems as a single light source, we showed that they can be described by orbital periods where a difference between projected velocities of components does not sufficiently degrade an observed spectrum that it no longer recognized as Solar-like. The spectroscopic signature and radial velocity variations were further used to put a limit on the minimum orbital period of an inner pair to be at least 20~years. Shorter periods are not completely excluded as it could happen that the spectrum was acquired in a specific orbital phase where the difference between projected orbital speeds is negligible. From the fact that analysed objects are spatially unresolvable for the \G\ spacecraft, the orbital size of outer binary pair can extend up to 100-350~AU and therefore have orbital periods of order of a few hundred years.

To confirm their existence, detected systems are ideal candidates to be observed with precise interferometric measurements or high time-resolution photometers if they happen to be occulted by the Moon. Simulation of the lunar motion showed that four of the analysed multiple candidates lie in its path if the observations would be carried-out from the Asiago observatory that has suitable photon counting detectors.

The main drawback of the analysis was found to be its separate treatment of photometric and spectroscopic information in two independent fitting procedures. In future analyses, they should be combined to acquire even more precise results as different stellar physical parameters have a different degree of impact on these two types of measurements.

\section*{Acknowledgments}
\label{sec:ack}
This work is based on data acquired through the Australian Astronomical Observatory, under programmes: A/2014A/25, A/2015A/19, A2017A/18 (The GALAH survey); A/2015A/03, A/2015B/19, A/2016A/22, A/2016B/12, A/2017A/14 (The K2-HERMES K2-follow-up program); A/2016B/10 (The HERMES-TESS program); A/2015B/01 (Accurate physical parameters of Kepler K2 planet search targets); S/2015A/012 (Planets in clusters with K2). We acknowledge the traditional owners of the land on which the AAT stands, the Gamilaraay people, and pay our respects to elders past and present.

K\v{C}, TZ, and JK acknowledge financial support of the Slovenian Research Agency (research core funding No. P1-0188 and project N1-0040).

This research was partly supported by the Australian Research Council 
Centre of Excellence for All Sky Astrophysics in 3 Dimensions (ASTRO 3D), 
through project number CE170100013.

YST is supported by the NASA Hubble Fellowship grant HST-HF2-51425.001 awarded by the Space Telescope Science Institute.

Follow-up observations were collected at the Copernico telescope (Asiago, Italy) of the INAF - Osservatorio Astronomico di Padova.

This work has made use of data from the European Space Agency (ESA) mission {\it Gaia} (\url{https://www.cosmos.esa.int/gaia}), processed by the {\it Gaia} Data Processing and Analysis Consortium (DPAC, \url{https://www.cosmos.esa.int/web/gaia/dpac/consortium}). Funding for the DPAC has been provided by national institutions, in particular the institutions participating in the {\it Gaia} Multilateral Agreement.

\bibliographystyle{mnras}
\bibliography{bib_trojnice}

\appendix

\section{Table description and summary}
\label{sec:outres}

In the Table \ref{tab:out_table} we provide a list of metadata available for every object detected using the methodology described in this paper. The complete table of detected objects and its metadata is available in electronic form at the CDS. An excerpt of the table, containing a subset of columns, for definitive and probable triple candidates is given in Table \ref{tab:out_triple}.

\begin{table}
\centering
\caption{List and description of the fields in the published catalogue of analysed objects.}
\label{tab:out_table}
\begin{tabular}{l c l}
	\hline
	Field & Unit & Description \\ 
	\hline
	\texttt{source\_id} & & {\it Gaia} DR2 source identifier \\
	\texttt{sobject\_id} & & Unique internal per-observation star ID \\
	\texttt{ra} & deg & Right ascension from {\it Gaia} DR2 \\
	\texttt{dec} & deg & Declination from {\it Gaia} DR2 \\
	\texttt{ruwe} & & Value of re-normalized astrometric $\chi^2$ \\
	
	\texttt{m\_sim\_p} & & Photometric $\chi^2$ for original parameters \\
	\texttt{m\_sim\_f} & & Spectroscopic $\chi^2$ for original parameters \\
	
	\texttt{s1\_teff1} & K & \Teffn{1} in a fitted single system \\
	\texttt{s1\_feh} & & \Feh\ of a fitted single system \\
	\texttt{s1\_sim\_p} & & Photometric $\chi^2$ of a fitted single system \\
	\texttt{s1\_sim\_f} & & Spectroscopic $\chi^2$ of a fitted single system\\

	\texttt{s2\_teff1} & K & \Teffn{1} in a fitted binary system \\	
	\texttt{s2\_teff2} & K & \Teffn{2} in a fitted binary system \\
	\texttt{s2\_feh} & & \Feh\ of a fitted binary system \\
	\texttt{s2\_sim\_p} & & Photometric $\chi^2$ of a fitted binary system \\
	\texttt{s2\_sim\_f} & & Spectroscopic $\chi^2$ of a fitted binary system\\
	
	\texttt{s3\_teff1} & K & \Teffn{1} in a fitted triple system \\	
	\texttt{s3\_teff2} & K & \Teffn{2} in a fitted triple system \\
	\texttt{s3\_teff3} & K & \Teffn{3} in a fitted triple system \\
	\texttt{s3\_feh} & & \Feh\ of a fitted triple system \\
	\texttt{s3\_sim\_p} & & Photometric $\chi^2$ of a fitted triple system \\
	\texttt{s3\_sim\_f} & & Spectroscopic $\chi^2$ of a fitted triple system\\
	
	\texttt{n\_stars\_p} & & Best fitting photometric configuration \\
	\texttt{n\_stars\_f} & & Best fitting spectroscopic configuration\\
	
	\texttt{class} & & Final configuration classification \\
    \texttt{flag} & & Result quality flags \\
	\hline
\end{tabular}
\end{table}

\begin{table*}
	\centering
	\caption{Subset of results for definitive and probable Solar-like triple candidates detected by our selection and fitting procedure. The complete table is given as a supplementary material to this paper in a form of the textual CSV file. It is also available in the electronic form at the CDS portal.}
	\label{tab:out_triple}
\begin{tabular}{ccccccccccc}
	\hline
	source\_id & ruwe & s2\_teff1 & s2\_teff2 & s2\_feh & s3\_teff1 & s3\_teff2 & s3\_teff3 & s3\_feh & class & flag \\
	\hline
	6157059919188478720 & 11.1 & 6002 & 6000 & 0.24 & 5825 & 5787 & 5780 & 0.03 & 3 & 1 \\
	6777339306532222080 & 8.0 & 5875 & 5819 & 0.10 & 5642 & 5636 & 5631 & -0.06 & 3 & 1 \\
	6412815502155127808 & 1.1 & 5880 & 5316 & -0.02 & 5922 & 4703 & 4701 & 0.03 & >2 & 4 \\
	6564302331580491904 & 1.1 & 5991 & 4701 & 0.13 & 5945 & 4702 & 4700 & 0.02 & >2 & 0 \\
	5386113598793714304 & 1.5 & 5882 & 5808 & 0.07 & 5878 & 5438 & 5313 & -0.03 & 3 & 5 \\
	2534579880633620992 & 1.2 & 5986 & 5876 & 0.15 & 5750 & 5739 & 5714 & -0.06 & 3 & 6 \\
	5484353352124904064 & 1.1 & 5876 & 5802 & 0.14 & 5844 & 5525 & 5425 & -0.00 & >2 & 0 \\
	5399712362903401984 & 1.3 & 5968 & 4703 & 0.08 & 5922 & 4703 & 4701 & 0.01 & >2 & 4 \\
	3688523450018482432 & 1.1 & 5820 & 5688 & -0.015 & 5998 & 4717 & 4702 & 0.09 & >2 & 2 \\
	6220408320279116032 & 7.6 & 5923 & 5883 & 0.11 & 5724 & 5677 & 5669 & -0.08 & 3 & 1 \\
	6198738457927949440 & 0.9 & 6052 & 4798 & 0.15 & 5986 & 4704 & 4701 & 0.10 & >2 & 4 \\
	5822222074090352384 & 4.7 & 5651 & 5640 & -0.04 & 5866 & 4702 & 4701 & 0.01 & >2 & 1 \\
	6355462192511955456 & 1.0 & 5700 & 5692 & -0.11 & 5471 & 5449 & 5379 & -0.28 & >2 & 4 \\
	4678229218054642176 & 0.9 & 6028 & 4791 & 0.09 & 5968 & 4702 & 4700 & 0.09 & >2 & 4 \\
	4423111085550775680 & 1.6 & 5829 & 5774 & -0.02 & 5645 & 5633 & 5398 & -0.11 & >2 & 1 \\
	6261736621608044416 & 4.7 & 5927 & 5890 & 0.07 & 6154 & 4702 & 4701 & 0.20 & >2 & 1 \\
	5947219160131475712 & 1.1 & 6071 & 5843 & 0.22 & 6007 & 5724 & 5265 & 0.10 & 3 & 4 \\
	6661888382295654656 & 1.5 & 5725 & 5678 & -0.07 & 5956 & 4706 & 4701 & 0.04 & >2 & 1 \\
	6362136502970853120 & 1.8 & 5844 & 5826 & 0.013 & 5666 & 5658 & 5540 & -0.18 & >2 & 1 \\
	6645693508028329728 & 1.3 & 5936 & 4715 & 0.06 & 5833 & 4703 & 4701 & -0.02 & >2 & 4 \\
	\hline
\end{tabular}
\end{table*}

\section{Additional figures}
In order to increase the readability and transparency of the text, additional and repeated plots are supplied as appendices to the main text.

\begin{figure}
	\centering
	\includegraphics[width=\columnwidth]{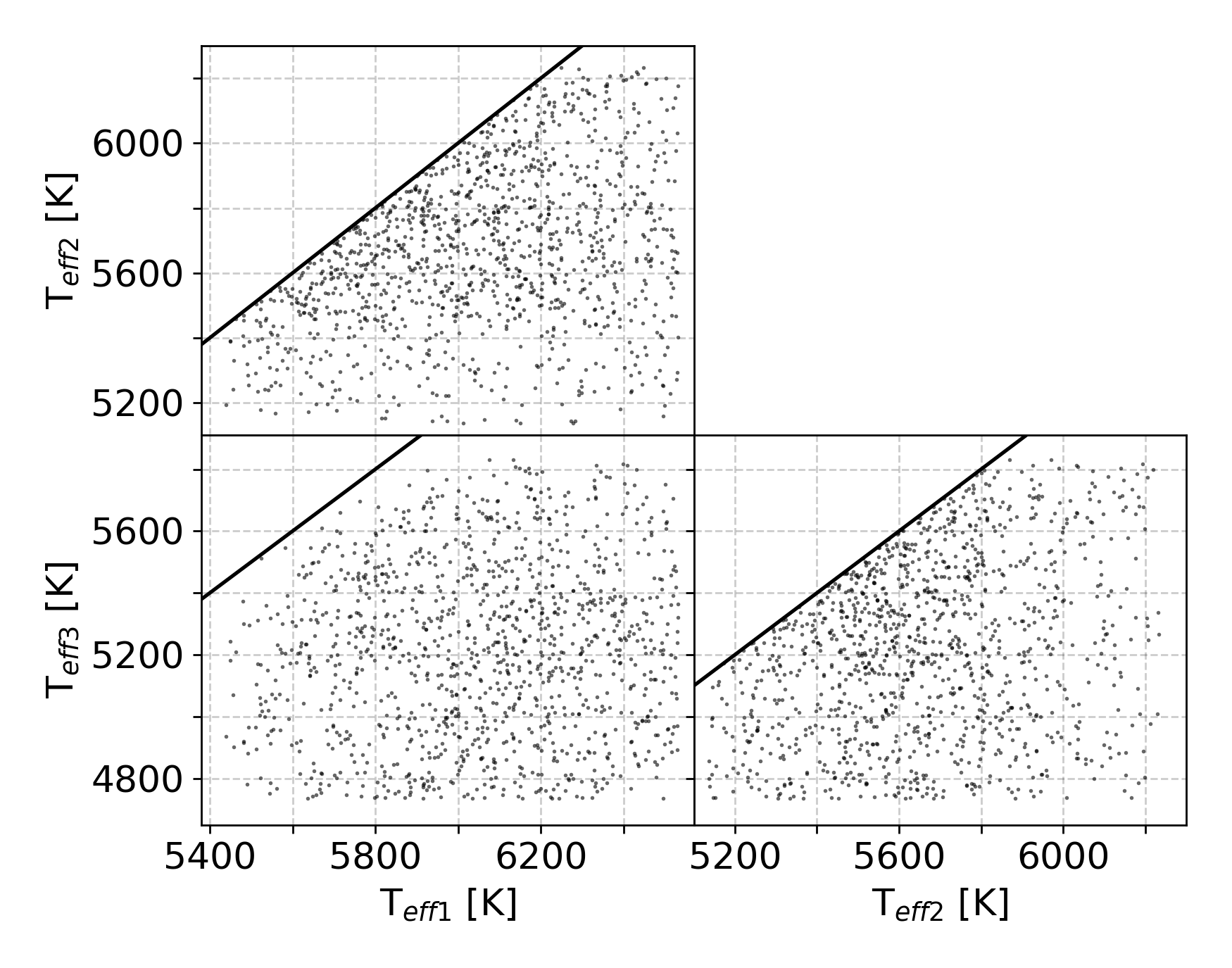}
	\caption{Initial distribution of walker parameters considered in the photometric fit. To ensure unique solutions with increasing \Teff, combinations above the linear line were not considered in the fit.}
	\label{fig:teff_initial_walkers}
\end{figure}

\begin{figure}
	\centering
	\includegraphics[width=\columnwidth]{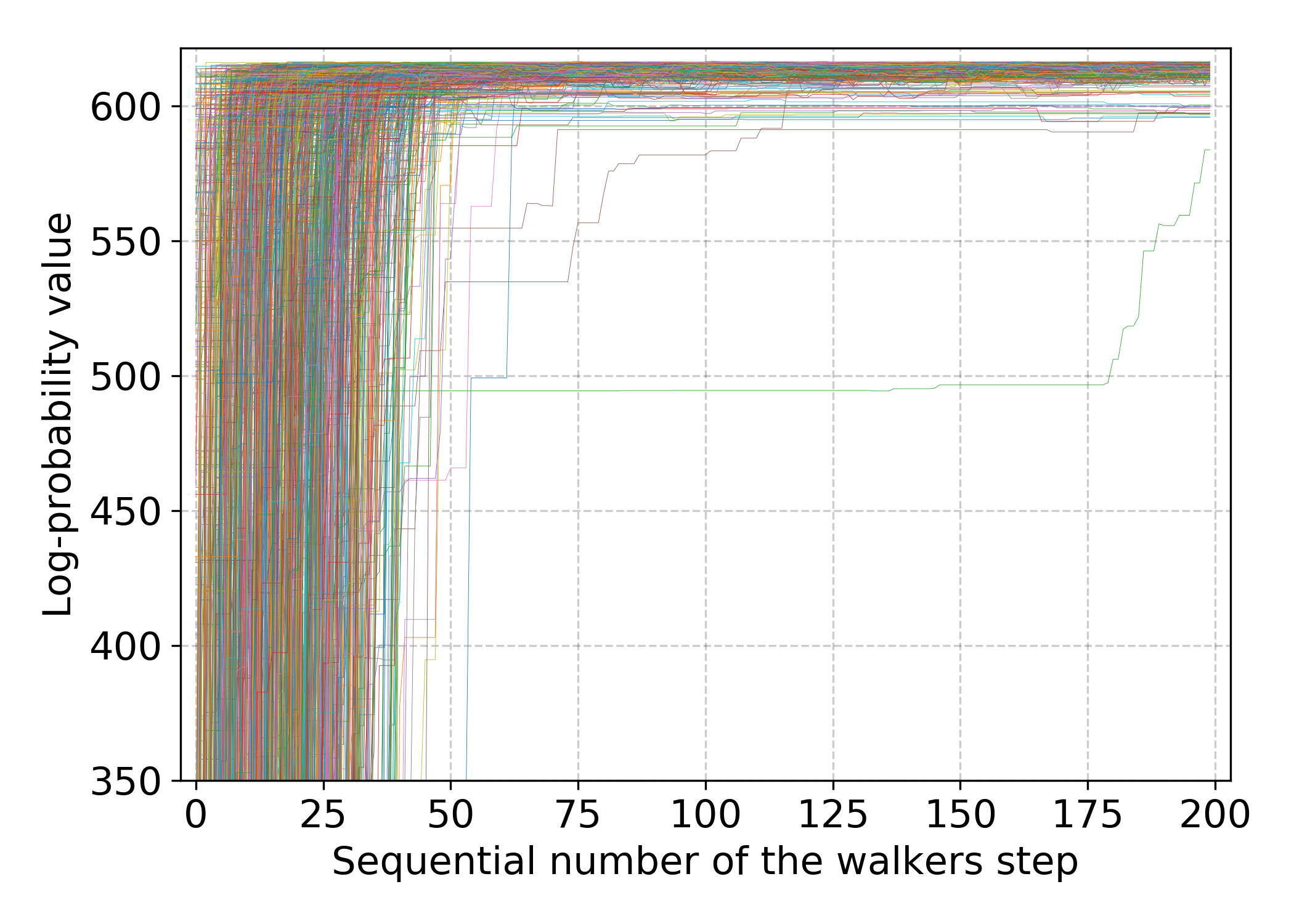}
	\caption{Convergence of walkers in the initial photometric fit. The plot shows a log-probability of the posteriors shown in the Figure \ref{fig:posterior_dist}. Walkers converge to the same value of log-probability after ~50 steps. Every walker is plotted with different colour.}
	\label{fig:walkers_logprob}
\end{figure}

\begin{figure}
	\centering
	\includegraphics[width=\columnwidth]{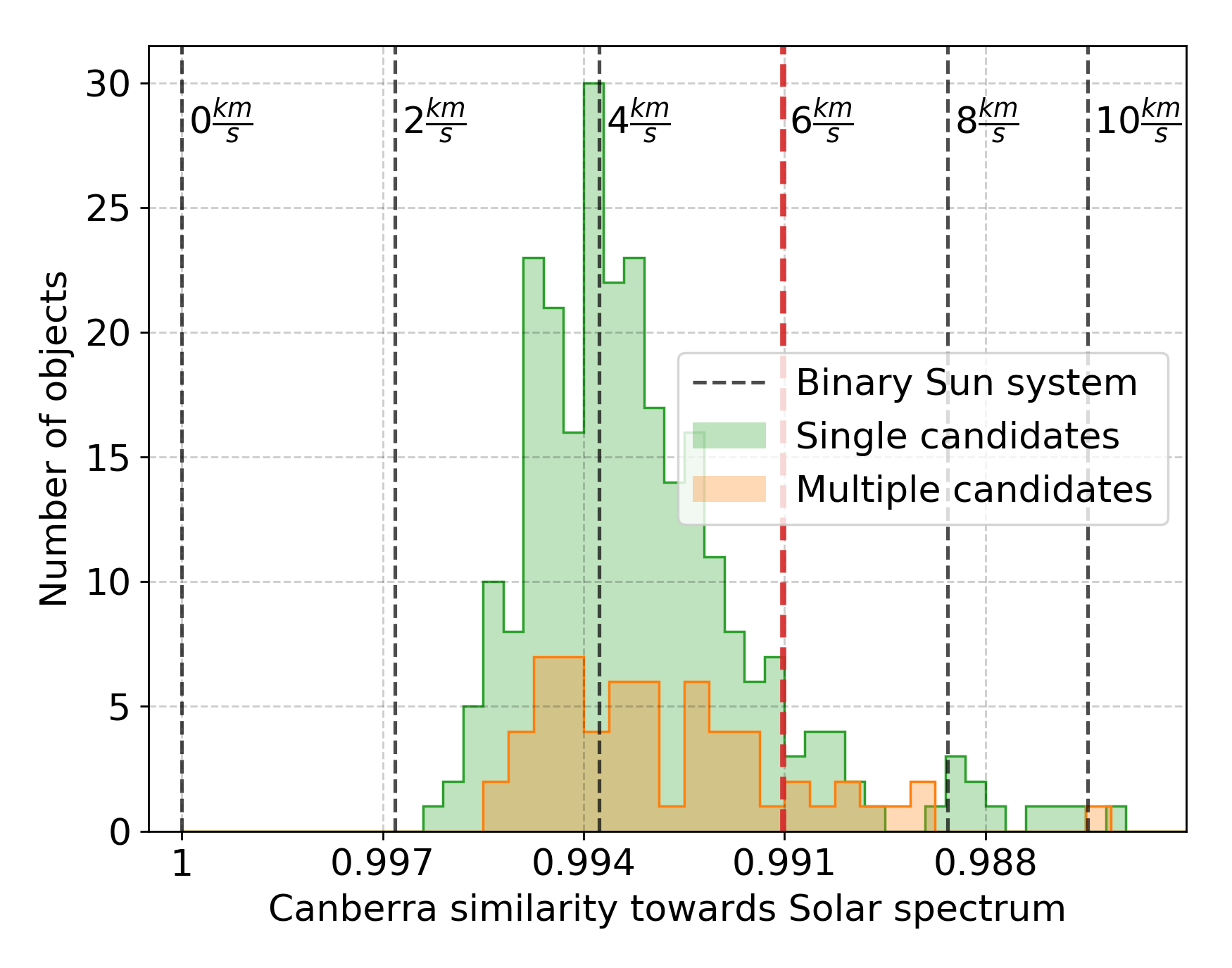}
	\caption{Same as Figure \ref{fig:rv_similarity} but for the case of a triple system where only one component out of three has a radial velocity shift in comparison to other two.}
	\label{fig:rv_similarity_2}
\end{figure}

\begin{figure}
	\centering
	\includegraphics[width=\columnwidth]{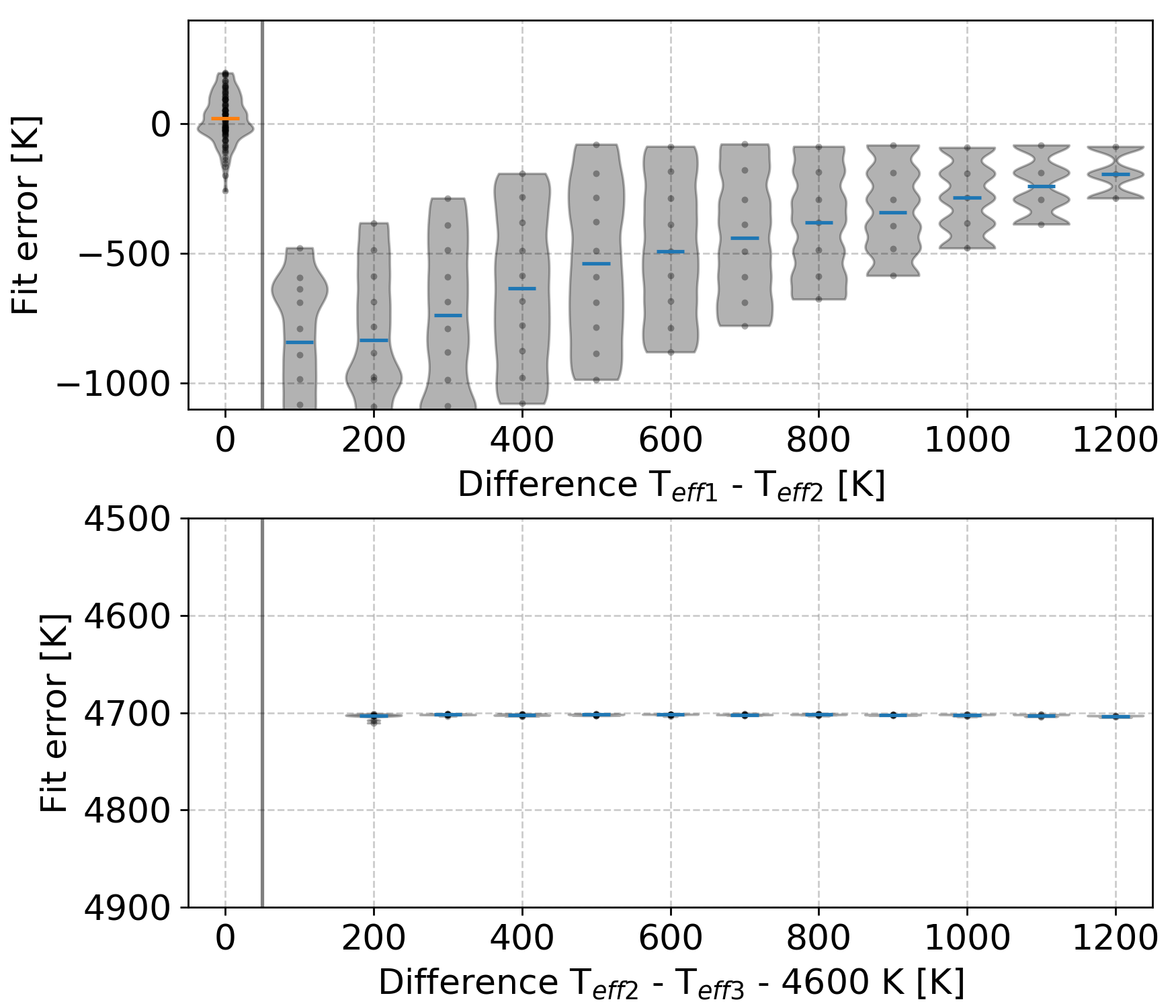}
	\caption{Similar to a plot in Figure \ref{fig:triple_sym_res}, but representing a case when we try to describe a synthetic binary system with \Teffn{3}~=~$0$ using a triple star model.}
	\label{fig:binary_sym_res}
\end{figure}

\begin{figure*}
	\centering
	\includegraphics[width=0.98\textwidth]{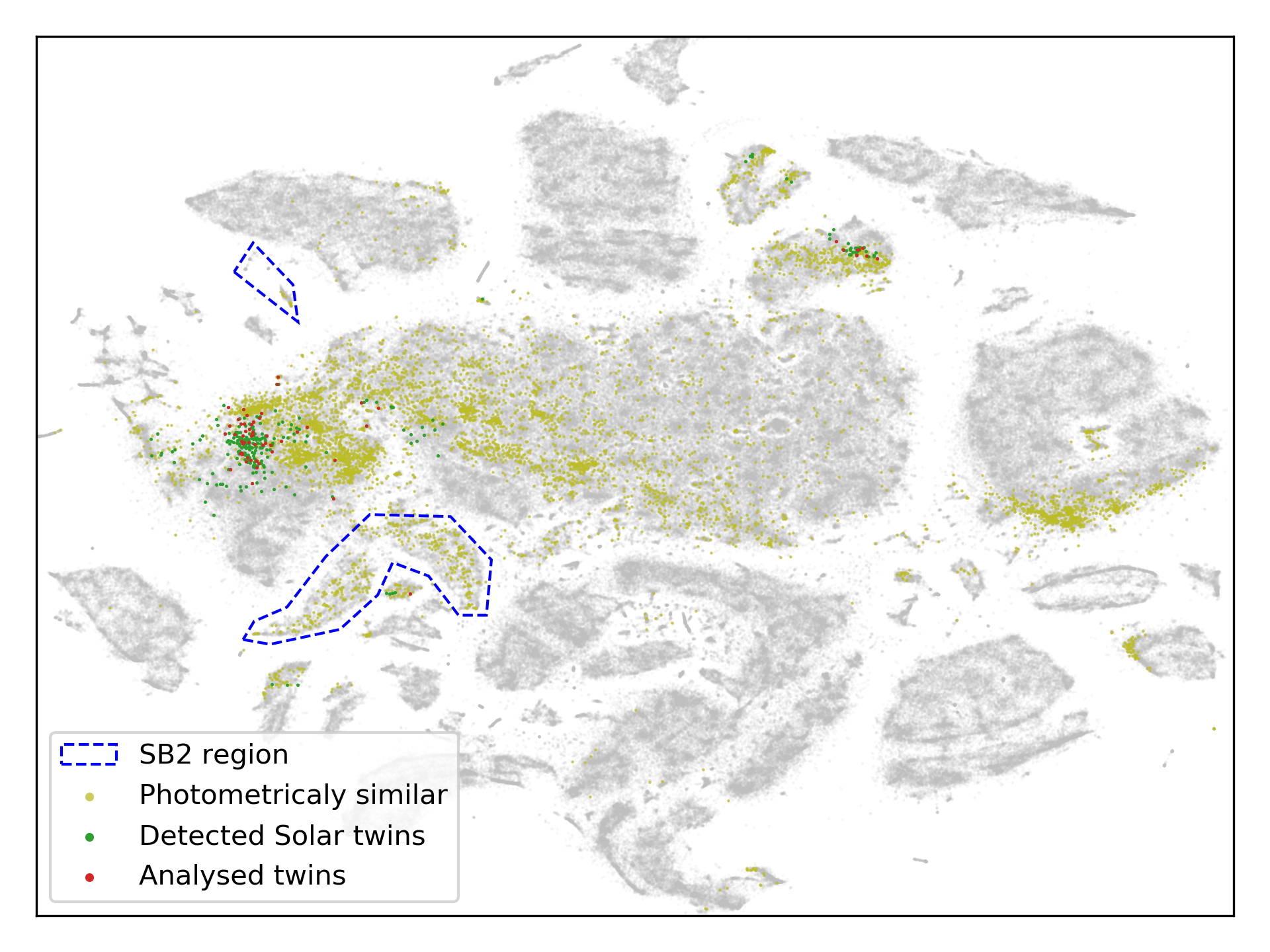}
	\caption{Visual representation of similarities between spectra using dimensionality reduction analysis. Clumps in this 2D projection represent morphologically similar spectra, whose features separate them from the rest of the data set. Blue-dashed polygons indicate over-densities, where SB2 spectra are located \citep[projection and regions are taken from Figure 13 in][]{buder2018}. Each grey dot represent one spectrum of GALAH survey. Green coloured dots represent objects considered in this paper, of which objects marked with red were analysed for higher order multiplicity. Yellowish dots represent objects whose \G\ magnitudes fall inside the range of determined triple star candidates.}
	\label{fig:tsne_marked}
\end{figure*}

\bsp	
\label{lastpage}
\end{document}